\documentclass[aps,pre,preprint,groupedaddress,showpacs]{revtex4}
\usepackage{graphicx}
\usepackage{amsmath}
\usepackage{amsfonts,amssymb}
\newcommand{\ket}[1]{| #1 \rangle}

\newcommand{\rb}[1]{\left( #1 \right)}

\newcommand{\go}{\rightarrow}
\newcommand{\ew}[1]{\langle{#1}\rangle}
\begin{document}

% Use the \preprint command to place your local institutional report
% number in the upper righthand corner of the title page in preprint mode.
% Multiple \preprint commands are allowed.
% Use the 'preprintnumbers' class option to override journal defaults
% to display numbers if necessary
%\preprint{}

%Title of paper
\title{Chaos and the Quantum Phase Transition in the Dicke Model}

% repeat the \author .. \affiliation  etc. as needed
% \email, \thanks, \homepage, \altaffiliation all apply to the current
% author. Explanatory text should go in the []'s, actual e-mail
% address or url should go in the {}'s for \email and \homepage.
% Please use the appropriate macro foreach each type of information

% \affiliation command applies to all authors since the last
% \affiliation command. The \affiliation command should follow the
% other information
% \affiliation can be followed by \email, \homepage, \thanks as well.
\author{Clive Emary}
\email[]{emary@theory.phy.umist.ac.uk}
\author{Tobias Brandes}
%\homepage[]{Your web page}
%\thanks{}
%\altaffiliation{}
\affiliation{ Department of Physics,
           UMIST,
           P.O. Box 88,
           Manchester 
           M60 1QD, 
	   U. K.}

%Collaboration name if desired (requires use of superscriptaddress
%option in \documentclass). \noaffiliation is required (may also be
%used with the \author command).
%\collaboration can be followed by \email, \homepage, \thanks as well.
%\collaboration{}
%\noaffiliation

\date{\today}

\begin{abstract}
 We investigate the quantum chaotic properties of the Dicke Hamiltonian; 
a quantum-optical model
which describes 
a single-mode bosonic 
field interacting with an ensemble of $N$ two-level atoms.
This model
exhibits a zero-temperature quantum phase transition in the $N \go \infty$
limit, which we describe exactly in an effective Hamiltonian approach.
We then numerically investigate the system 
at finite $N$ and, 
by analysing the level statistics, we 
demonstrate that the system undergoes a transition from
quasi-integrability to quantum chaotic, and that this 
transition is caused by the precursors of the quantum phase-transition.
Our considerations of the wavefunction indicate that this is 
connected with a delocalisation of the system and the emergence 
of macroscopic coherence.  We also derive a semi-classical Dicke 
model, which exhibits analogues of
all the important features of the quantum model, such as the phase 
transition and the concurrent onset of chaos.

% LocalWords:  Dicke bosonic eigen integrable observables
% LocalWords:  integrability wavefunction delocalisation

\end{abstract}

% insert suggested PACS numbers in braces on next line
% insert suggested keywords - APS authors don't need to do this
%\keywords{}

%\maketitle must follow title, authors, abstract, \pacs, and \keywords
\maketitle

%\newpage
\section{Introduction}
   
Chaos plays a key role in considerations concerning the 
boundary between the classical 
and quantum worlds, 
not just because of the importance of chaos in 
classical physics \cite{clchaos}, 
but becaus there is no direct analogue
of chaos in quantum mechanics \cite{gutz}.
The linearity of quantum
dynamics precludes the characteristic
exponential sensitivity to initial conditions
of classical chaos, and forces us to look for what 
have become known as ``signatures of quantum chaos'' -  properties
whose presence in the quantum system would lead us to expect the 
corresponding classical motion to be chaotic \cite{ha:ak}.  
Several such signatures
have been identified, such as level statistics \cite{be:ta,bo:gs}, 
level dynamics \cite{lvl:dyn}, 
and sensitivity to initial perturbation \cite{Loschmidt}.

An oft encountered feature of quantum-chaotic systems is
that as some parameter is varied, these signatures bespeak 
a cross-over from integrable to quantum chaotic behaviour.  
This parameter may, for example, describe the character 
of boundary conditions, such as the shape of 
a quantum billiard 
\cite{bo:gs},
the distribution of random fluctuations in disorder models
\cite{zh:kr,ja:sh,gesh:shards,gesh:Qcomp},
or the strength of some non-linear potential or interaction 
\cite{de:ga,zmkc,mp:bs,ibm,za:ku,jfywh}.
A large class of model may be described by
a Hamiltonian of the form
\begin{eqnarray}
  H = H_0 + \lambda V,
\label{HH0V}
\end{eqnarray}
where, although $H_0$ is integrable, the full Hamiltonian $H$ is not for
any $\lambda \ne 0$.  Here, increasing the parameter $\lambda$ from zero
upwards gradually drives the system away from 
integrability and towards chaos.  A well studied, albeit time-dependent,
example is the kicked rotator \cite{ha:ak}, where the parameter $\lambda$
is the kick-strength.

In this paper, we consider a system of the type described by Hamiltonian
(\ref{HH0V}), but unlike the typically one-dimensional or
non-interacting models, we shall consider a system of
$N$ {\em interacting} particles, 
in a situation where many-body and collective effects are 
critical.  Specifically, the model we study exhibits a
quantum phase transition (i.e. one at zero temperature \cite{sachdev})
in the thermodynamic limit 
of $N \go \infty$ at a critical value of the parameter, $\lambda_c$.

The influence of a quantum phase transition (QPT) on the
transition to chaos has been studied in but a handful of cases.
Important examples include the three-dimensional Anderson 
model, where the metal-insulator transition is accompanied
by a change in the level-statistics \cite{zh:kr}, and models
of spin glass shards \cite{gesh:shards}, 
which have found topical application in 
the study of the effects of quantum chaos on quantum computing
\cite{gesh:Qcomp}.  Heiss and co-workers have 
investigated the connection between the onset of chaos near a 
QPT and the exceptional points of the spectrum \cite{Heiss:excep},
both generically  and for the specific example of the Lipkin 
model \cite{Heiss:QPT}.

In order to investigate the impact of QPT on the signatures of 
quantum chaos, we study the Dicke 
Hamiltonian (DH) \cite{di:ck}, which is of key importance as a model 
describing collective effects in quantum optics 
\cite{andreev,benedict}.  We demonstrate that there is a clear connection
between the precursors of the QPT and the onset of quantum chaos 
as manifested in the level-statistics.  We are able to understand this
connection by studying the wavefunctions of the system, and by deriving a
semi-classical analogue of this intrinsically quantum system.  The
current publication is an extension of our previous work \cite{ce:tb1}.

In the form considered here, the DH describes a collection of $N$ 
two-level atoms interacting with a single bosonic mode via a 
dipole interaction with an atom-field coupling strength $\lambda$.
The DH may be written
\begin{eqnarray}
H = \hbar \omega_0 J_z + \hbar\omega a^\dagger a 
     + \frac{\lambda}{\sqrt{2j}} \rb{a^\dagger + a}\rb{J_+ + J_-}, 
\label{DHamID}
\end{eqnarray}
where $a$, $a^\dagger$ describe a bosonic mode of frequency $\omega$,
and the angular momentum operators $\left\{J_i;~~i=z,\pm \right\}$ 
describe the ensemble of
two-level atoms of level-splitting $\omega_0$ 
in terms of a pseudo-spin of length $j=N/2$.
The thermodynamic limit of $N\go \infty$ is thus equivalent to making
the length of the pseudo-spin tend to infinity $j \go \infty$.  
The DH is usually considered 
in the standard quantum optics approach of the rotating-wave 
approximation (RWA), which is valid for small values of the coupling 
$\lambda$, and involves neglecting the counter-rotating 
terms $a^\dag J_+$ and $a J_-$.
This makes the DH integrable, simplifying the analysis but 
also removing the possibility of quantum chaos.
Dicke used this model to illustrate the importance of collective 
effects in the atom-light interaction \cite{di:ck}, leading to the concept of 
super-radiance, where the atomic ensemble spontaneously emits 
with an intensity proportional to $N^2$ rather than $N$, as one 
would expect if the atoms were radiating incoherently \cite{benedict}.

The phase transition in the DH was first described by Hepp and Lieb 
\cite{he:li}, and a mathematically more transparent treatment was provided
by Wang and  Hioe \cite{wa:ho}.  
They considered the thermodynamics of the model 
in the RWA
and concluded that for a coupling of 
$\lambda < \sqrt{\omega \omega_0}$, no phase 
transition occurs for any temperature, whereas for 
$\lambda > \sqrt{\omega \omega_0}$, there exists a critical 
temperature $T_c$ given by
\begin{eqnarray}
  \frac{1}{k_B T_c}
  = \frac{2\omega}{\omega_0}\mathrm{artanh}
  \rb{\frac{\omega \omega_0}{\lambda^2}},
\end{eqnarray}
at which point the system undergoes a phase transition.  
Above the critical temperature,
the system is in the effectively unexcited ``normal phase'', 
whereas for $T<T_c$ the system is in the ``super-radiant phase'', 
a macroscopically excited 
and highly collective
state which possesses 
the potential to super-radiate.

In contrast to this earlier work, 
we shall consider this phase transition 
at zero temperature, where increasing
the coupling $\lambda$ through a critical value 
of $\lambda_c=\sqrt{\omega \omega_0}/2$ drives the system to undergo a 
transition from the normal to the super-radiant phase
(the difference between this critical coupling $\lambda_c$ 
and the value quoted for the finite-temperature case arises because
the latter has been derived in the RWA, 
which renormalises the critical coupling by a factor of 
two \cite{he:li2,cr:du}).  Here, we derive exact results 
without the RWA for 
the energy spectrum and eigenfunctions 
in the thermodynamic limit 
by employing a bosonisation technique based upon the 
Holstein-Primakoff transformation of the angular 
momentum algebra \cite{ho:pr,ho:pr2}.  This enables us to derive 
an effective Hamiltonian 
to describe the system in each of its two phases. 
One important step that we make is the introduction of an abstract 
position-momentum representation for both the field and atomic 
systems.  This not only facilitates the formulation of the exact 
solutions, but also provides us with a useful way of visualising 
the wavefunctions across the phase transition.
There is a discrete ``parity''
symmetry associated with this model, and at the 
phase-transition this symmetry becomes broken.
This QPT has been discussed in the RWA by Hillery and Mlodinow 
\cite{hi:ll}, using an effective Hamiltonian method 
that is similar to ours.  
However, having illustrated the existence of the QPT, 
they concentrated solely on the normal phase, and
were not interested in chaos.

Away from the thermodynamic limit at finite $N$ and $j$, 
the DH is, in general, non-integrable.
Quantum-chaotic properties of the DH have been discussed by 
several authors 
\cite{fu:ru,kus,mil,gr:ho1,gr:ho2,le:ne,msln,fingea} but, to
the best of our knowledge, have never been connected with the QPT, 
and a systematic study 
of the dependence of the systems behaviour on the number of atoms $N$
is lacking.
Graham and H\"ohnerbach  have contributed extensively 
to the discussion \cite{gr:ho1}, 
especially in relation to the special case of 
spin-$1/2$ (the Rabi Hamiltonian), and have outlined many 
semi-classical and approximate schemes for these 
systems.  Moreover, they have provided a preliminary analysis of 
the level statistics of the DH, concluding that spectra 
of the type associated with quantum chaos do occur 
for certain, isolated parameter values \cite{gr:ho2}.  
Several authors have conducted studies of chaos in various 
(semi-)classical models related to DH
\cite{fu:ru,mil,msln,fingea}.  That there have been 
several different semi-classical models is a consequence of the
ambiguity in describing quantum spins in classical terms.
The influence of the QPT also seems to have have been overlooked in these
semi-classical models.

We consider the quantum-mechanical system away from 
the thermodynamic limit by using numerical diagonalisation, 
and examine the energy 
spectra of the system for signatures of 
quantum chaos.  We consider 
the nearest-neighbour level-spacing distribution function $P\rb{S}$,
which is perhaps the best-known signature of quantum chaos \cite{ha:ak}.
We calculate the $P\rb{S}$ for various values of $N$ and $\lambda$ 
and demonstrate a clear connection between the change in $P\rb{S}$ from
quasi-integrable to quantum chaotic and the coupling at which 
the QPT occurs, $\lambda_c$.  We then proceed to consider the 
wavefunctions of the system at finite $N$
using an abstract position-momentum representation. This 
enables us to conclude that the precursors of the QPT give rise to a 
localisation-delocalisation transition in which the ground-state 
wavefunction bifurcates into a macroscopic superposition for 
any $N < \infty$. 

As mentioned above, there has been much work in trying to find 
a semi-classical analogue of the DH \cite{fu:ru,mil,msln,fingea}.
The bosonisation procedure that we employ here allows us to write 
the DH in terms of a pair of coupled harmonic oscillators.  This 
suggests a very natural semi-classical analogue of the DH, obtained by 
simply replacing the quantum oscillators with classical ones.
We demonstrate that our semi-classical model 
reflects the quantum behaviour better than 
those of previous studies.
Specifically, our semi-classical model exhibits a symmetry-breaking phase
transition in the limit that $N \go \infty$, and we show that 
the precursors of 
this classical transition give rise to the onset 
of classical chaos, in close agreement with the quantum model.  An analogue
of the macroscopic superposition is also evident.  In our conclusions, 
we pay special attention to the meaning of a classical limit for 
the DH, and in particular the relevance of the 
semi-classical model derived here.

The paper is organised as follows.  In section 
\ref{model} we introduce the DH fully.
Exact solutions are derived in the thermodynamic limit in section \ref{TDL}.
Section \ref{Ochaos} sees an analysis of the level-statistics and 
wavefunctions of the system at finite $j$.  Our semi-classical model 
is derived in section \ref{clas}, and its phase transition and 
chaotic properties discussed.  We discuss briefly the 
differences between the full DH and the Hamiltonian in RWA in section
\ref{RWA}, before we draw our final conclusions in section \ref{disc}.
Some of our exact expressions are reproduced in the Appendix.
% LocalWords:  integrable Poissonian linearity integrability towards QPT Dicke
% LocalWords:  DH bosonic boson RWA Hepp Lieb Hioe ohnerbach Rabi Dyson rotator
% LocalWords:  bosonisation Primakoff observables macroscopically visualising
% LocalWords:  wavefunctions diagonalisation spacings localisation wavefunction
% LocalWords:  delocalisation superposition behaviour eigenfunctions organised
% LocalWords:  billiard Heiss co Lipkin QPTs unexcited renormalises Hillery
% LocalWords:  Mlodinow neighbour

\newpage
\section{The Dicke Hamiltonian \label{model}}
  
The full Dicke Hamiltonian (DH) models the interaction of $N$ atoms 
with a number of bosonic field modes via 
dipole interactions within an ideal cavity \cite{di:ck}.  
We initially represent the atoms as a collection of $N$ identical, 
but distinguishable 
two-level systems each with level-splitting $\omega_0$.  The $i$th 
atom is described by the spin-half operators
$\left\{ s_k^{(i)};~k=z,\pm\right\}$, obeying the commutation rules
$\left[s_z,s_\pm \right]=\pm s_\pm$; $\left[s_+,s_- \right]= 2 s_z$.
These two-level atoms interact 
with $M$ bosonic modes, which have frequencies 
$\left\{ \omega_\alpha\right\}$,
interact with coupling strengths $\left\{\lambda_\alpha\right\}$,
and are described by the bosonic creation and annihilation operators 
$\left\{a^\dag_\alpha\right\}$ and $\left\{a_\alpha\right\}$. 
In terms of these quantities the full DH is given by
\begin{eqnarray}
  H = \omega_0 \sum_{i=1}^N s_z^{(i)}
    + \sum_{\alpha=1}^M \omega_\alpha a^\dagger_\alpha a_\alpha
    + \sum_{\alpha=1}^M \sum_{i=1}^N
        \frac{\lambda_\alpha}{\sqrt{N}} \rb{a^\dagger_\alpha + a_\alpha}
        \rb{s^{(i)}_+ + s^{(i)}_-},
  \label{fullDH}
\end{eqnarray}
where we have set $\hbar =1$.
The origin of the factor $1/ \sqrt{N}$ in the interaction is
the fact that the original dipole coupling 
strength is proportional to
$1/ \sqrt{V}$, where $V$ is the volume of the cavity.
By writing $\rho = N/V$, where $\rho$ is 
the density of the atoms in the cavity, this becomes
$\sqrt{\rho / N}$ and by subsuming the density into
the coupling constants, $\left\{ \lambda_\alpha\right \}$, we obtain
$1/ \sqrt{N}$ explicitly in the coupling.

In Eq. (\ref{fullDH}) we have not made the usual rotating-wave 
approximation 
(RWA) under which one would neglect the counter-rotating terms
$a^\dagger_\alpha s_+^{(i)}$ 
and $a_\alpha s_-^{(i)}$.  We shall consider aspects of of the
RWA in section \ref{RWA}.

We now specialise the Hamiltonian 
to consider a single mode bosonic field, 
and thus we drop the subscript $\alpha$.
The analysis of this Hamiltonian is further simplified by the 
introduction of collective atomic operators,
\begin{eqnarray}
  J_z \equiv \sum_{i=1}^N s_z^{(i)};
  ~~~ J_\pm \equiv \sum_{i=1}^N s_\pm^{(i)}.
\end{eqnarray}
These operators obey the usual 
angular momentum commutation relations,
\begin{equation}
\left[J_z,J_\pm \right]=\pm J_\pm
;~~~~~~
\left[J_+,J_- \right]= 2 J_z.
\end{equation}
The Hilbert space of this algebra is spanned by the kets 
$\left\{\ket{j,m};~m=-j,-j+1,\ldots,j-1,j\right\}$, which are 
known as the Dicke 
states, and are eigenstates of $\bold{J}^2$ and $J_z$:   
$J_z \ket{j,m} = m\ket{j,m}$ and $\mathbf{J}^2 \ket{j,m} = j\rb{j+1}\ket{j,m}$.
The raising and lowering operators act on these states in the following
way: $J_\pm \ket{j,m}= \sqrt{j\rb{j+1}-m\rb{m\pm 1}}~ \ket{j,m\pm 1}$.
Note that
$j$ corresponds to Dicke's ``co-operation number'' which
takes the values $\frac{1}{2},\frac{3}{2},\ldots,\frac{N}{2}$
for $N$ odd, and $0,1,\ldots,\frac{N}{2}$ for $N$ even.
For example, with $N=2$ atoms, $j$ can take the values $0$ and $1$.
In terms of the $s_z$ values of the individual spins, 
the sector with $j=1$ contains the 
triplet states $\ket{\downarrow \downarrow}$, 
$\frac{1}{\sqrt{2}}\rb{\ket{\uparrow \downarrow}+\ket{\downarrow \uparrow}}$ 
and $\ket{\uparrow \uparrow}$.  
The $j=0$ sector contains only the singlet state,
$\frac{1}{\sqrt{2}}\rb{\ket{\uparrow \downarrow}-\ket{\downarrow \uparrow}}$.
In general, the set of 
atomic configurations for $N>2$ is non-trivial \cite{arrechi}, and
in terms of the individual atom configurations, the
states are non-separable and contain 
entanglement \cite{DSentangle}.
In this work, we shall take $j$ to have its maximal value, $j=N/2$, 
and once set, this value of $j$ is constant, as the interaction in the DH
does not mix $j$-sectors.
Thus, the collection of $N$ two-level systems is described as a 
single $\rb{N+1}$-level system, which is viewed as a large 
pseudo-spin vector of length $j=N/2$.

In terms of the collective operators, the single-mode DH
may be written
\begin{eqnarray}
H = \omega_0 J_z + \omega a^\dagger a 
     + \frac{\lambda}{\sqrt{2j}} \rb{a^\dagger + a}\rb{J_+ + J_-}.
\label{DHam1}
\end{eqnarray}
In the following, when we refer to the Dicke Hamiltonian 
we shall mean this single-mode Hamiltonian unless otherwise stated.
The resonance condition is $\omega = \omega_0$, and when plotting 
results we generally work on scaled resonance, such that 
$\omega = \omega_0 =1$.

Associated with the DH is a conserved parity $\Pi$,
such that $\left[H,\Pi\right]=0$, which is given by
\begin{equation}
  \Pi= \exp\left\{i\pi \hat{N}\right\};~~~
  \hat{N} = a^\dagger a + J_z+j,
  \label{parity}
\end{equation}
where $\hat{N}$ is the ``excitation number'' 
and counts the total number of excitation quanta in the system.
$\Pi$ possesses two eigenvalues, $\pm 1$, depending on
whether the number of quanta is even or odd, and correspondingly the 
Hilbert-space of the total system is split into two non-interacting 
sub-spaces.  

\begin{figure}[tb]
  \centerline{\includegraphics[clip=true,width=0.5\textwidth]{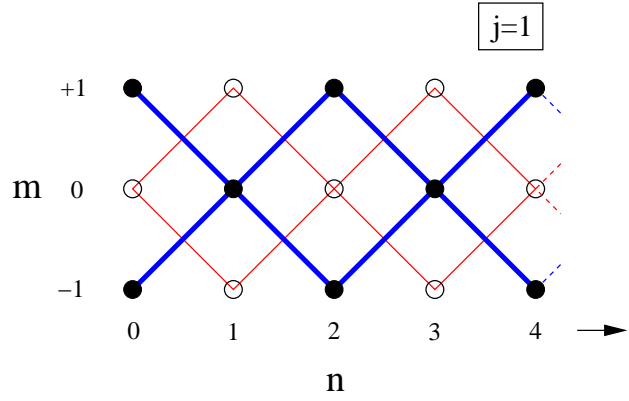}}
  \caption{\label{lattice}
    Schematic lattice representation of the states of the Dicke model for
    the example of $j=1$.
    Shaded (unshaded) dots denote states of positive (negative) parity,
    with solid lines representing the couplings between the states.
  }
\end{figure}

If we express the Hilbert-space of the total system in terms of the 
basis $\left\{\ket{n}\otimes \ket{j,m}\right\}$, 
where $\ket{n}$ are number states of the field, 
$a^\dagger a \ket{n} = n \ket{n}$, and $\ket{j,m}$ are 
the Dicke states, the DH and the significance of the parity operator
may be viewed in a simple lattice analogy.  We construct a 
two-dimensional lattice, each point of which represents 
a basis vector and is labeled
$\rb{n,m}$.  An example of this lattice with $j=1$ is shown in Fig. 
\ref{lattice}.  Note that the lattice is finite in the `$m$' direction, 
but infinite in the `$n$' direction, reflecting the dimensionality of the 
Hilbert-space.
In this picture, 
we see that because the interaction conserves the
parity $\Pi$, states with an even total excitation number $n + m +j$ 
interact only with other even states, and odd states interact only with odd 
states.  This has the effect of dividing the total lattice into 
two inter-weaved sub-lattices, 
which correspond to the two different parity sectors.

%%%%%%%%%%%%%%%%%%%%%%%%%%%%%%%%%%%%%%%%%%%%%%%%%%%%%%%%%%%%%%%%%%%
%%%%%%%%%%%%%%%%%%%%%%%%%%%%%%%%%%%%%%%%%%%%%%%%%%%%%%%%%%%%%%%%%%

% LocalWords:  Dicke bosonic th RWA boson kets eigenstates Dicke's co DH Eq
% LocalWords:  integrability unshaded singlet dimensionality specialise

\newpage
\section{Thermodynamic Limit \label{TDL}}
  
We begin by considering the DH in the thermodynamic limit, in which the
number of atoms becomes infinite, $N \go \infty$,
and hence $j \go \infty$.  
In this limit, the DH undergoes a QPT 
at a critical value of the atom-field coupling strength
$\lambda_c=\sqrt{\omega \omega_0}/2$, at which point the symmetry
associated with the parity operator $\Pi$ of Eq. (\ref{parity})
is broken.  To 
describe this QPT we shall derive two effective Hamiltonians, one 
to describe the system in the normal phase $\lambda<\lambda_c$, and one 
to describe it in the broken-symmetry, super-radiant phase 
$\lambda>\lambda_c$.  It should be noted that the results 
derived below are exact in 
this limit, and this allows us to understand the nature of this 
system in a very detailed way.

In this analysis we shall make extensive use of the Holstein-Primakoff
representation of the angular momentum operators, 
which represents the operators in terms of a single bosonic mode 
in the following way \cite{ho:pr,ho:pr2},
\begin{eqnarray}
  J_+ = b^\dagger \sqrt{2j - b^\dagger b} &;&~~~
  J_- = \sqrt{2j - b^\dagger b}~ b
  \nonumber \\
  J_z &=& \rb{b^\dagger b - j}, \label{HP1}
\end{eqnarray}
where the introduced Bose operators obey $\left[b,b^\dag\right]=1$.

Making these substitutions into the DH of Eq. (\ref{DHam1}),
we obtain the two-mode bosonic Hamiltonian
\begin{eqnarray}
  H = \omega_0 \rb{b^\dagger b - j} + \omega a^\dagger a
  + \lambda \rb{a^\dagger + a}
  \rb{
    b^\dagger \sqrt{1-\frac{b^\dagger b}{2j}}  
    + \sqrt{1-\frac{b^\dagger b}{2j}}b 
  }.   \label{DHam2}
\end{eqnarray}
In this representation the parity operator $\Pi$ becomes
\begin{eqnarray}
  \Pi= \exp\left\{i\pi \left[a^\dagger a + b^\dagger b \right]\right\},
  \label{altpar}
\end{eqnarray}
and the analogy with the standard parity operator of a two-dimensional 
harmonic operator is thus apparent \cite{bi:v4}.

\subsection{Normal phase}

We derive an effective Hamiltonian for the system 
in the normal phase by simply neglecting 
terms in the full Hamiltonian of Eq. (\ref{DHam2}) with $j$ in the 
denominator.  This approximates the square-root in the 
Holstein-Primakoff mapping with unity,
and we obtain the effective Hamiltonian $H^{(1)}$ given by
\begin{equation}
  H^{(1)} = \omega_0 b^\dagger b + \omega a^\dagger a
   + \lambda \rb{a^\dagger + a}\rb{b^\dagger + b} - j \omega_0,
\label{lcDHjinf}
\end{equation}
which is bi-linear in the bosonic operators and can thus 
be simply diagonalised.  This is most easily facilitated by the 
introduction of position and momentum operators
for the two bosonic modes,
\begin{eqnarray}
x = \frac{1}{\sqrt{2 \omega}}\rb{a^\dagger + a};& 
p_x =i\sqrt{\frac{\omega}{2}}\rb{a^\dagger - a} \nonumber\\
y =\frac{1}{\sqrt{2 \omega_0}}\rb{b^\dagger + b} ;& 
p_y = i\sqrt{\frac{\omega_0}{2}}\rb{b^\dagger - b}.\label{x-ycoord}
\end{eqnarray}
This representation will be particularly useful when we
come to  consider the 
wavefunctions of the system.  Expressing Hamiltonian $H^{(1)}$ 
in terms of these operators we obtain
\begin{equation}
H^{(1)} = \frac{1}{2}
\left\{ \omega^2 x^2 + p_x^2 + \omega_0^2 y^2 + p_y^2 
+ 4 \lambda \sqrt{\omega \omega_0}~ x y - \omega_0 - \omega\right\}
-j \omega_0 ,
\label{H1jinf}
\end{equation}
which may be diagonalised by rotating the coordinate system in 
the following way
\begin{equation}
x = q_1 \cos \gamma^{(1)} + q_2 \sin \gamma^{(1)};~~~
y = - q_1 \sin \gamma^{(1)} + q_2 \cos \gamma^{(1)}, \label{lcrot}
\end{equation}
where the angle $\gamma^{(1)}$ is given by
\begin{equation}
\tan \rb{2\gamma^{(1)}} = \frac{4 \lambda \sqrt{\omega \omega_0}}
{\omega_0^2 - \omega^2}.
\end{equation}
On resonance, $\omega=\omega_0$, $\gamma^{(1)} = \pi/4$, so that
$x=\rb{q_1+q_2}/\sqrt{2}$ and $y=\rb{-q_1+q_2}/\sqrt{2}$.
This rotation eliminates the $xy$ interaction term in 
the Hamiltonian, which then assumes the form of two 
uncoupled oscillators,
\begin{equation}
H^{(1)} = \frac{1}{2} 
\left\{ {\varepsilon^{(1)}_-}^2 q_1^2 + p_1^2 
+ {\varepsilon^{(1)}_+}^2 q_2^2 + p_2^2 
- \omega -\omega_0 \right\} - j \omega_0.
\label{lcDH2}
\end{equation}
We now re-quantise $H^{(1)}$ with the introduction
of two new bosonic modes defined by
\begin{eqnarray}
q_1 = \frac{1}{\sqrt{2 \varepsilon^{(1)}_-}}\rb{c_1^\dagger + c_1};& 
p_1 =i\sqrt{\frac{\varepsilon^{(1)}_-}{2}}\rb{c_1^\dagger - c_1} \nonumber\\
q_2 =\frac{1}{\sqrt{2 \varepsilon^{(1)}_+}}\rb{c_2^\dagger + c_2} ;& 
p_2 = i\sqrt{\frac{\varepsilon^{(1)}_+}{2}}\rb{c_2^\dagger - c_2},
\end{eqnarray}
and arrive at the final diagonal form
\begin{equation}
H^{(1)}  =  \varepsilon^{(1)}_- c_1^\dagger c_1 
+ \varepsilon^{(1)}_+  c^\dagger_2 c_2 
+\frac{1}{2} \rb{\varepsilon^{(1)}_+ + \varepsilon^{(1)}_- -\omega - \omega_0}
 - j \omega_0.
\end{equation}
The bosonic operators 
$\left\{c_1,c_1^\dag, c_2,c_2^\dag \right\}$, in terms of which
$H^{(1)}$ is diagonal,
are linear combinations of the original operators
$\left\{a,a^\dag, b,b^\dag \right\}$, as detailed in 
appendix \ref{appBOGT}, and describe collective atom-field 
excitations.  The energies of the two independent 
oscillator modes
$\varepsilon_\pm^{(1)}$ are given by
\begin{equation}
{\varepsilon^{(1)}_{\pm}}^2 = \frac{1}{2}\left\{ \omega^2 + \omega_0^2 
\pm \sqrt{\rb{\omega_0^2 - \omega^2}^2 
+ 16 \lambda^2 \omega \omega_0}\right\}\label{lcepspm}.
\end{equation}
Crucially, we see that the excitation energy $\varepsilon_-^{(1)}$ is
real only when
$ \omega^2 + \omega_0^2 \ge  \sqrt{\rb{\omega_0^2 - \omega^2}^2 
+ 16 \lambda^2 \omega \omega_0} $, or equivalently
$\lambda \le \sqrt{\omega \omega_0}/2=\lambda_c$.
Thus we see that $H^{(1)}$ remains valid for $\lambda \le \lambda_c$,
i.e. in the normal phase.
In this phase, the ground-state energy
is given by $E^{(1)}_G  = -j\omega_0 $, which is ${\cal O}\rb{j}$, 
whereas the excitation
energies $\varepsilon_\pm^{(1)}$ are ${\cal O}\rb{1}$.  This means that
scaling our energies with $j$, 
the excitation spectrum above the ground state
becomes quasi-continuous in the $j\go \infty$ limit, that is to say that the 
excitation energies differ by an infinitesimal amount from $E_G$.

It should be noted that $H^{(1)}$ commutes with the parity operator 
$\Pi$, and thus the eigenstates of $H^{(1)}$ have definite parity,
with the 
ground state having positive parity.
This can been seen from the fact that at $\lambda = 0$, the 
ground-state is $\ket{0}\ket{j,-j}$ in the original $\ket{n}\ket{j,m}$ 
basis, which clearly has an even excitation number, $n+m+j=0$.
As the energy levels in the normal phase 
are non-degenerate, the continuity of the ground state with 
increasing $\lambda$ ensures that it always has positive parity in 
this phase.

\subsection{Super-radiant phase }

In order to describe the system above the phase transition, we must 
incorporate the fact that both the field and the atomic ensemble
acquire macroscopic occupations.  To do this, we start with
the Holstein-Primakoff transformed Hamiltonian 
of Eq. (\ref{DHam2}) and 
displace the bosonic modes in either of the following ways
\begin{eqnarray}
a^\dagger \rightarrow c^\dagger + \sqrt{\alpha};
~~~b^\dagger \rightarrow d^\dagger - \sqrt{\beta}.
\label{disp1}
\end{eqnarray}
or
\begin{eqnarray}
a^\dagger \rightarrow c^\dagger - \sqrt{\alpha};
~~~b^\dagger \rightarrow d^\dagger + \sqrt{\beta}.
\label{disp2}
\end{eqnarray}
Crucially, we assume that the as yet undetermined parameters 
$\alpha$ and $\beta$ are of the  ${\cal O}\rb{j}$,
equivalent to assuming that both modes acquire
non-zero, macroscopic mean-fields above $\lambda_c$.
In the following we shall just consider the displacements 
given by Eq. (\ref{disp1}), as the calculation with the other choice is
identical but for a few changes of sign.

Making these displacements, the Hamiltonian  of Eq. (\ref{DHam2}) becomes
\begin{eqnarray}
H = \omega_0 \left\{d^\dagger d 
           - \sqrt{\beta}\rb{d^\dagger + d} + \beta -j\right\}
+ \omega \left\{c^\dagger c + \sqrt{\alpha}\rb{c^\dagger + c} + \alpha\right\}
\nonumber\\
+ \lambda\sqrt{\frac{k}{2j}} \rb{c^\dagger + c + 2\sqrt{\alpha}}
  \rb{d^\dagger \sqrt{\xi} + \sqrt{\xi}d - 2\sqrt{\beta} \sqrt{\xi}}, 
\label{hcDH2}
\end{eqnarray}
where for brevity we have written $
\sqrt{\xi} \equiv \sqrt{1 - \frac{d^\dagger d
- \sqrt{\beta}\rb{d^\dagger + d}}{k}}$ and 
$k \equiv 2j-\beta$.  Taking the thermodynamic limit 
by expanding the square-root $\sqrt{\xi}$ and then
setting terms with overall powers of $j$ in 
the denominator to zero, we obtain
\begin{eqnarray}
H^{(2)} &=& \omega c^\dagger c
+\left\{
   \omega_0 + \frac{2 \lambda}{k} \sqrt{\frac{\alpha \beta k}{2j}}
 \right\} d^\dagger d
- \left\{
   2 \lambda \sqrt{\frac{\beta k}{2j}} - \omega \sqrt{\alpha}
 \right\} \rb{c^\dagger + c} \nonumber \\
&&+\left\{
   \frac{4\lambda}{k}\sqrt{\frac{\alpha k}{2j}}\rb{j-\beta} 
  - \omega_0 \sqrt{\beta}
 \right\} \rb{d^\dagger + d}
+ \frac{\lambda}{2k^2}\sqrt{\frac{\alpha \beta k}{2j}} \rb{2k + \beta}
\rb{d^\dagger + d}^2 \nonumber \\
&&+\frac{2\lambda}{k} \sqrt{\frac{k}{2j}}\rb{j-\beta}
\rb{c^\dagger + c}\rb{d^\dagger + d} \nonumber \\
&&+\left\{
   \omega_0 \rb{\beta-j} + \omega \alpha 
   - \frac{\lambda}{k} \sqrt{\frac{\alpha \beta k}{2j}} \rb{1+4k}
 \right\}.
\label{hcDHhorror}
\end{eqnarray}
We now eliminate the terms in the $H^{(2)}$ that are linear in the
bosonic operators by choosing the displacements $\alpha$ and $\beta$
so that
\begin{eqnarray}
2\lambda\sqrt{\frac{ \beta k}{2j}} - \omega\sqrt{\alpha} =0,
\end{eqnarray}
and
\begin{eqnarray}
\left\{ \frac{4\lambda^2}{\omega j} \rb{j -\beta} - \omega_0\right\} 
\sqrt{\beta} =0.
\end{eqnarray}
The 
$\sqrt{\beta} = \sqrt{\alpha} =0$ solution of these equations  
recovers the normal phase Hamiltonian $H^{(1)}$.  The non-trivial 
solution gives
\begin{eqnarray}
\sqrt{\alpha} = \frac{2 \lambda}{\omega}\sqrt{\frac{j}{2}\rb{1 - \mu^2}}
~~~;&~~~&
\sqrt{\beta} = \sqrt{j \rb{1- \mu}},
\label{abdet}
\end{eqnarray}
where we have defined
\begin{eqnarray}
\mu \equiv \frac{\omega \omega_0}{4 \lambda^2} 
             = \frac{\lambda_c^2}{\lambda^2} .
\end{eqnarray}
With these determinations, the effective 
Hamiltonian of Eq. (\ref{hcDHhorror}) becomes
\begin{eqnarray}
H^{(2)} &=& \omega c^\dagger c
+ \frac{\omega_0}{2\mu}\rb{1+\mu} d^\dagger d
+ \frac{\omega_0 \rb{1-\mu}\rb{3+\mu}}{8 \mu \rb{1+\mu}} \rb{d^\dagger + d}^2
\nonumber \\
&&+ \lambda \mu \sqrt{\frac{2}{1+\mu}}\rb{c^\dagger + c}\rb{d^\dagger + d}
-j\left\{
    \frac{2 \lambda^2}{\omega}+ \frac{\omega_0^2 \omega}{8 \lambda^2}  
  \right\}
- \frac{\lambda^2}{\omega}\rb{1-\mu}.\label{hcDHab}
\end{eqnarray}
To facilitate the diagonalisation of this bi-linear Hamiltonian 
we move to a position-momentum representation defined by
\begin{eqnarray}
X \equiv \frac{1}{\sqrt{2 \omega}}\rb{c^\dagger + c};& 
P_{X} \equiv i\sqrt{\frac{\omega}{2}}\rb{c^\dagger - c} 
\nonumber\\
Y \equiv\frac{1}{\sqrt{2 \widetilde{\omega}}}\rb{d^\dagger + d} ;& 
P_{Y} \equiv i\sqrt{\frac{\widetilde{\omega}}{2}}
  \rb{d^\dagger - d},\label{XY}
\end{eqnarray}
where $\tilde{\omega} =  \frac{\omega_0}{2\mu}\rb{1+\mu}$.  
Note that this is not the same representation as defined in 
Eq. (\ref{x-ycoord}).  The diagonalisation then proceeds similarly
to before, involving a rotation in the $X$-$Y$ plane to the new coordinates
\begin{eqnarray}
X = Q_{1} \cos\gamma^{(2)} + Q_{2}\sin\gamma^{(2)}\nonumber\\
Y = -Q_{1}\sin\gamma^{(2)} + Q_{2}\cos\gamma^{(2)}
\label{XYQ1Q2}
\end{eqnarray}
with the angle $\gamma^{(2)}$ is given by
\begin{eqnarray}
\tan \rb{2 \gamma^{(2)}} = \frac{2\omega\omega_0\mu^2}
				{\omega_0^2 - \mu^2 \omega^2}.
\end{eqnarray}
Subsequent requantisation in terms of two new modes, 
$e^{(2)}_\pm$, corresponding to the rotated, decoupled 
oscillators gives us the diagonal form
\begin{eqnarray}
H^{(2)} &=& \varepsilon^{(2)}_- e_{1}^\dagger e_{1} 
+ \varepsilon^{(2)}_+  e^\dagger_{2} e_{2}
-j\left\{
    \frac{2 \lambda^2}{\omega}+ \frac{\omega_0^2 \omega}{8 \lambda^2}  
  \right\}\nonumber \\
&+&\frac{1}{2} 
\rb{
  \varepsilon^{(2)}_+ + \varepsilon^{(2)}_- 
  -\frac{\omega_0}{2 \mu}\rb{1+\mu} - \omega 
  -\frac{2\lambda^2}{\omega}\rb{1-\mu}
 },\label{hcDH}
\end{eqnarray}
with the oscillator energies being given by
\begin{eqnarray}
2{\varepsilon_\pm^{(2)}}^2 =
  \frac{\omega_0^2}{\mu^2} +  \omega^2
  \pm \sqrt{\left[ \frac{\omega_0^2}{\mu^2} - \omega^2 \right]^2 
  + 4\omega^2 \omega_0^2}.
\end{eqnarray}
The Bogoliubov transformations that induce this diagonalisation
are given in appendix \ref{appBOGT}.  The excitation energy
$\varepsilon_-^{(2)}$, and hence $H^{(2)}$,
remains real provided that 
$ \frac{\omega_0^2}{\mu^2} +  \omega^2  
  \ge\sqrt{\left[ \frac{\omega_0^2}{\mu^2} - \omega^2 \right]^2 
  + 4\omega^2 \omega_0^2}$, or equivalently
$\lambda \ge \sqrt{\omega \omega_0}/2 = \lambda_c$.  Thus we see that
$H^{(2)}$  describes the system in the super-radiant phase,
$\lambda \ge \lambda_c$, in which the
scaled ground-state energy is given by 
$E^{(2)}_G /j = -\left\{
\frac{2 \lambda^2}{\omega}+ \frac{\omega_0^2 \omega}{8 \lambda^2}  
\right\}$.

If we choose the signs of the operator displacements as per 
Eq. (\ref{disp2}), we obtain exactly the same values of $\alpha$
and $\beta$, and an effective Hamiltonian identical in form with 
Eq. (\ref{hcDH}), This clearly has the same spectrum and therefore,
each and every 
level of the total spectrum is doubly degenerate above the 
phase transition.
What has occurred is that the symmetry of the ground state, defined 
by the operator $\Pi$, has become spontaneously broken at $\lambda_c$.  
The Hamiltonian $H^{(2)}$, for either choice of displacement,
does not commute with $\Pi$, and thus its eigenfunctions do not
possess good parity symmetry.  

Although the global symmetry $\Pi$ becomes broken at the phase 
transition, two new local symmetries appear, corresponding to 
to the operator
\begin{eqnarray}
  \Pi^{(2)} \equiv  \exp\left\{i\pi 
  \left[c^\dagger c +d^\dagger d \right]\right\},
  \label{Pi2}
\end{eqnarray}
for both the two different choices of mean-field displacements.  This 
operator commutes with the appropriate super-radiant Hamiltonian,
$\left[H^{(2)},\Pi^{(2)}\right]=0$.

%%%%%%%%%%%%%%%%%%%%%%%%%%%%%%%%%%%%%%%%%%%%%%%%%%%%%%%%%%%%%%%%%%%%%%%%%%
%%%%%%%%%%%%%%%%%%%%%%%%%%%%%%%%%%%%%%%%%%%%%%%%%%%%%%%%%%%%%%%%%%%%%%%%%%
\subsection{Phase Transition}

\begin{figure}[tb]
  \centerline{\includegraphics[clip=true,width=0.5\textwidth]
  {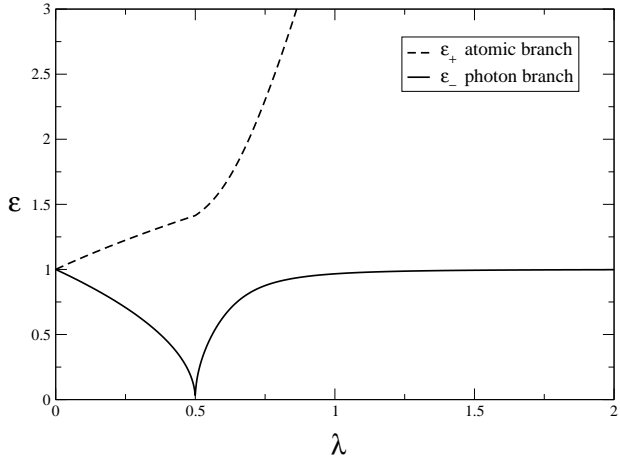}}
  \caption{\label{oscen1}
    The excitation energies of the 
    Dicke Hamiltonian in the thermodynamic limit as a 
    function of coupling $\lambda$.
    The Hamiltonian is resonant, $\omega = \omega_0 = 1$, and the vanishing
    of $\varepsilon_-$ at $\lambda=\lambda_c=0.5$ signals the occurrence
    of the QPT. }
\end{figure}

Having derived the two effective Hamiltonians which describe the 
system for all $\lambda$ in the $j \go \infty$ limit, 
we now describe the 
systems properties in each of its two phases.
The fundamental excitations of the system are given 
by the energies
$\varepsilon_\pm$, which
describe collective modes, similar to 
polariton modes in solid-state physics \cite{ea:li}.
The behaviour of these energies as a function of coupling
strength is displayed in Fig. \ref{oscen1},
%%%%%%%%%%%%%%%%%%%%%%%%%%%%%%%%%%%%%%%%%%%%%%%%%%%%%%%%%%%%%%%%%%%
%%%%%%%%%%%%%%%%%%%%%%%%%%%%%%%%%%%%%%%%%%%%%%%%%%%%%%%%%%%%%%%%%%
where we have labeled the two branches as ``atomic'' and 
``photonic'', according
to the nature of the excitation at zero coupling.  From this figure we
see that as the coupling approaches the critical value $\lambda_c$, 
the excitation energy of the photonic mode vanishes,
$\varepsilon_- \go 0$, as $\lambda \go \lambda_c$, demonstrating 
the existence of the QPT.  In contrast,
$\varepsilon_+$ tends towards a value of $ \sqrt{\omega_0^2 + \omega^2}$ 
as $\lambda \go \lambda_c$ from either direction.
In the asymptotic limit of $\lambda \go \infty$, $\varepsilon_- \go \omega$
(returning to its $\lambda=0$ value) whereas 
$\varepsilon_+ \go 4\lambda^2/\omega$.
The critical exponents of this QPT
are manifested in the behaviour of the
excitation energies \cite{sachdev}.  
As $\lambda \go \lambda_c$ from either direction, the energy 
$\varepsilon_-$ can be shown to vanish as
\begin{eqnarray}
\varepsilon_-\rb{\lambda \go \lambda_c} 
\sim 
\sqrt{
  \frac{32 \lambda_c^3 \omega^2}
       {16 \lambda_c^4 + \omega^4} 
}
 |\lambda_c - \lambda|^{1/2}.
\label{oscdiv}
\end{eqnarray}
The vanishing of $\varepsilon_-$ at $\lambda_c$ reveals this to be 
a second-order phase transition.
We define the characteristic length scale in the 
system in terms of this energy as 
\begin{eqnarray}
l_- = 1/\sqrt{\varepsilon_-}.
\end{eqnarray}
From Eq. (\ref{oscdiv}) this length diverges as
$|\lambda-\lambda_c|^{-\nu}$
with the exponent $\nu = 2$.  We then write 
that $\varepsilon_-$ vanishes as 
$|\lambda-\lambda_c|^{z\nu}$, with the dynamical critical 
exponent being given by $z=2$.
At the phase transition point, we have
\begin{eqnarray}
  H^{(1)}\rb{\lambda_c} =
  H^{(2)}\rb{\lambda_c} =
  \sqrt{\omega^2 + \omega^2} c_2^\dag c_2 + \frac{1}{2}
  \rb{\sqrt{\omega^2 + \omega^2}-\omega -\omega_0}-j\omega_0,
\end{eqnarray}
from which we see that at $\lambda_c$ the system becomes
effectively one-dimensional.

%%%%%%%%%%%%%%%%%%%%%%%%%%%%%%%%%%%%%%%%%%%%%%%%%%%%%%%%%%%%%%%%%%%%%
%%%%%%%%%%%%%%%%%%%%%%%%%%%%%%%%%%%%%%%%%%%%%%%%%%%%%%%%%%%%%%%%%%%%%
\begin{figure}[tb]
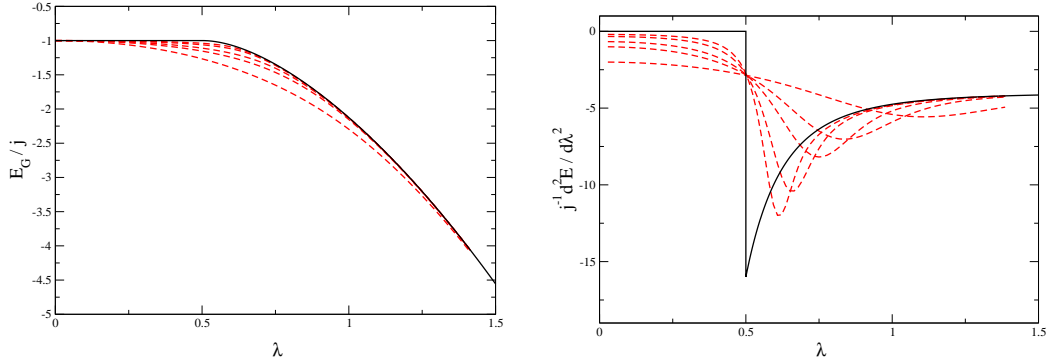

  \centerline{
    \includegraphics[clip=true,width=0.4\textwidth]{./Eg1.eps}
    ~~~
    \includegraphics[clip=true,width=0.4\textwidth]{./d2Edl2.eps}
             }
  \caption{\label{gse}
    The scaled ground-state energy $E_G/j$ and its second derivative 
    $j^{-1} d^2 E_G / d \lambda^2$ 
    as a 
    function of coupling $\lambda$.
    Solid lines denote results in the thermodynamic limit, whereas
    dashed lines correspond to the results for 
    various finite values of $j=\frac{1}{2},1,\frac{3}{2},3,5$.
    The Hamiltonian is resonant: $\omega = \omega_0 = 1$, 
    $\lambda_c=0.5$. 
   }
\end{figure}
%%%%%%%%%%%%%%%%%%%%%%%%%%%%%%%%%%%%%%%%%%%%%%%%%%%%%%%%%%%%%%%%%%%%%
%%%%%%%%%%%%%%%%%%%%%%%%%%%%%%%%%%%%%%%%%%%%%%%%%%%%%%%%%%%%%%%%%%%%%

The ground-state energy of the system $E_G$ is shown in Fig. \ref{gse}
and the analytic form expression is given in Table \ref{gsproptab}.
Note that we scale all quantities by $j$, which means 
that the plotted $E_G/j$ is equal to $2 E_G/N$, twice the
energy per atom.  We also plot the second derivative of the 
ground-state energy with respect to $\lambda$, which
possesses a discontinuity at $\lambda_c$, clearly locating the phase 
transition.

\begin{figure}[tb]
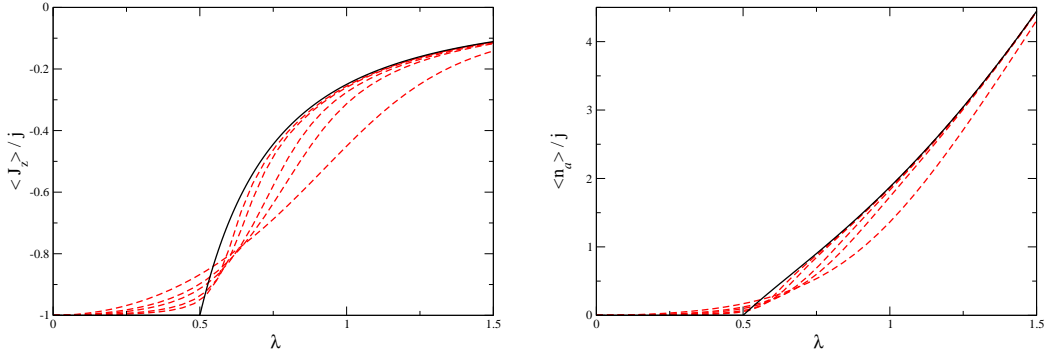

  \centerline{
    \includegraphics[clip=true,width=0.4\textwidth]{./ai1.eps}
     ~~~
    \includegraphics[clip=true,width=0.4\textwidth]{./pno1.eps}}
  \caption{\label{aipno}
    The scaled atomic inversion and mean photon number
    of the Dicke Hamiltonian as a 
    function of coupling $\lambda$.
    Solid lines denote results in the thermodynamic limit, whereas
    dashed lines correspond to the results for 
    various finite values of $j=\frac{1}{2},1,\frac{3}{2},3,5$.
    The Hamiltonian is resonant: $\omega = \omega_0 = 1$, 
    $\lambda_c=0.5$. 
   }
\end{figure}

%%%%%%%%%%%%%%%%%%%%%%%%%%%%%%%%%%%%%%%%%%%%%%%%%%%%%%%%%%%%%%%%%%%%%
\begin{table}[bt]
  \begin{center}
   \begin{tabular}{|c|c|c|}
   \hline
   & $\lambda < \lambda_c$ 
     & $\lambda > \lambda_c$ \\
   \hline
   $E_G / j$ 
     & $-\omega_0$ 
     & $-\frac{2\lambda^2}{\omega} -  \frac{2\lambda_c^4}{\lambda^2 \omega}$ \\
   $\ew{J_z} / j$ 
     & $-1$ 
     & $-\lambda_c^2 / \lambda^2$ \\
   $\ew{a^\dagger a} / j$ 
     & $0$ 
     & ~$2\rb{\lambda^4 - \lambda_c^4} / \rb{\omega \lambda}^2$~\\
   \hline
  \end{tabular}
 \end{center}
 \caption{
   \label{gsproptab}
   The ground-state energy, atomic inversion and mean photon number of
   the Dicke Hamiltonian in the thermodynamic limit.
 }
\end{table}
%%%%%%%%%%%%%%%%%%%%%%%%%%%%%%%%%%%%%%%%%%%%%%%%%%%%%%%%%%%%%%%%%%%%%
%%%%%%%%%%%%%%%%%%%%%%%%%%%%%%%%%%%%%%%%%%%%%%%%%%%%%%%%%%%%%%%%%%%%%
In Fig. \ref{aipno} we plot the atomic inversion $\ew{J_z}/j$ and 
the mean photon number $\ew{n_a}/j\equiv\ew{a^\dag a}/j$.
This figure clearly illustrates the nature of the phase transition,
-- in the normal phase, the system is only microscpically excited,
whereas
above $\lambda_c$ both the field and atomic ensemble acquire 
macroscopic excitations.
We may write the values of the atomic inversion and the mean 
photon number above $\lambda_c$ in the following fashion:
\begin{eqnarray}
\ew{J_z}/j = 1 - \beta/j~,~~~\ew{a^\dag a}/j = \alpha/j~;~~~ \lambda>\lambda_c.
\end{eqnarray}
Thus making clear the physical meaning of the displacement parameters $\alpha$ 
and $\beta$ of Eqns. (\ref{abdet}).
%%%%%%%%%%%%%%%%%%%%%%%%%%%%%%%%%%%%%%%%%%%%%%%%%%%%%%%%%%%%%%%%%%%%%
%%%%%%%%%%%%%%%%%%%%%%%%%%%%%%%%%%%%%%%%%%%%%%%%%%%%%%%%%%%%%%%%%%%%%
%%%%%%%%%%%%%%%%%%%%%%%%%%%%%%%%%%%%%%%%%%%%%%%%%%%%%%%%%%%%%%%%%%%%%
%%%%%%%%%%%%%%%%%%%%%%%%%%%%%%%%%%%%%%%%%%%%%%%%%%%%%%%%%%%%%%%%%%%%%
%%%%%%%%%%%%%%%%%%%%%%%%%%%%%%%%%%%%%%%%%%%%%%%%%%%%%%%%%%%%%%%%%%%%%

%%%%%%%%%%%%%%%%%%%%%%%%%%%%%%%%%%%%%%%%%%%%%%%%%%%%%%%%%%%%%%%%%%%%%
\subsection{Ground-state Wavefunction}

We now consider the ground-state wavefunctions of the system above 
and below the phase transition.
After diagonalisation, the two effective Hamiltonians are both
of the form of a pair of uncoupled harmonic oscillators.  Thus, 
in the representation in 
which the Hamiltonians are diagonal, 
their wavefunctions will simply be the product of the appropriate 
harmonic oscillator 
eigenfunctions.  Here we seek to express these wavefunctions in terms of 
the two-dimensional $x$-$y$ representation of Eq. (\ref{x-ycoord}) - 
which corresponds to the original atomic and field degrees of freedom.

We have already noted that in the Holstein-Primakoff representation
the parity operator has the form 
$\Pi = \exp\left\{i\pi \left[a^\dagger a + b^\dagger b\right]\right\}$. 
From
our knowledge of the harmonic oscillator \cite{bi:v4}, 
we know that
the action of $\Pi$ in the $x$-$y$ representation is to perform
the coordinate inversions, $x \go -x$ and $y\go -y$, with $p_x$ and $p_y$ 
remaining unaffected.  Thus the operation of $\Pi$
is equivalent to a rotation of $\pi$ about the 
coordinate origin and, in the normal phase where $\Pi$ is a good quantum 
number, the wavefunctions will be seen to be 
invariant under this rotation.

The ground-state wavefunction of a single harmonic oscillator in terms of 
its coordinate $q$ is a Gaussian with  width 
determined by the energy of the oscillator.  Correspondingly we define 
the normalised Gaussian functions
\begin{eqnarray}
  G_\pm^{(1,2)}\rb{q} =
  \rb{\frac{\varepsilon^{(1,2)}_\pm}{\pi}}^{1/4}
  \exp\left\{-\frac{\varepsilon^{(1,2)}_\pm}{2}q^2 \right\},
\end{eqnarray}
where $\varepsilon^{(1,2)}_\pm$ are the excitation energies 
encountered earlier.

\begin{figure}[tb]
  \centerline{
  \includegraphics[clip=true,width=0.8\textwidth]{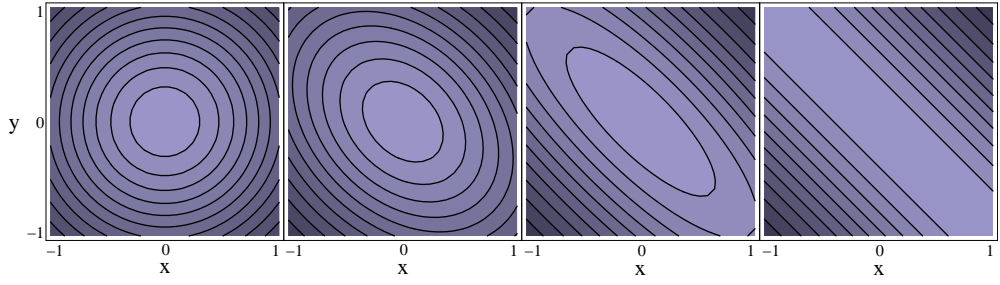}}
  \caption{\label{LCwavefn}
    The ground-state wavefunction $\Psi_G^{(1)}$ of the 
    low-coupling Hamiltonian $H^{(1)}$
    in the $x$-$y$ position-momentum representation 
    for couplings $\lambda = 0,0.3,0.49,0.4999999$.
    The Hamiltonian is resonant: $\omega = \omega_0 = 1$, 
    $\lambda_c=0.5$. 
  }
\end{figure}

In the normal phase, the effective Hamiltonian $H^{(1)}$ is 
diagonal in the $q_1$-$q_2$ representation of Eq. (\ref{lcrot}) and
its ground-state wavefunction $\Psi^{(1)}_G$
in this representation is therefore
\begin{eqnarray}
  \Psi^{(1)}_G\rb{q_1,q_2} = G_-^{(1)}\rb{q_1}G_+^{(1)}\rb{q_2}.
\end{eqnarray}
Moving to the $x$-$y$ representation, we have
\begin{eqnarray}
  \Psi^{(1)}_G\rb{x,y} = 
  G_-^{(1)}\rb{ x\cos \gamma^{(1)} - y\sin \gamma^{(1)}}
  G_+^{(1)}\rb{x\sin \gamma^{(1)} + y\cos \gamma^{(1)}},
\end{eqnarray}
and this wavefunction is plotted for various couplings in 
Fig. \ref{LCwavefn}.
%%%%%%%%%%%%%%%%%%%%%%%%%%%%%%%%%%%%%%%%%%%%%%%%%%%%%%%%%%%%%%%%%%
%%%%%%%%%%%%%%%%%%%%%%%%%%%%%%%%%%%%%%%%%%%%%%%%%%%%%%%%%%%%%%%%%%
At $\lambda = 0$ the wavefunction is the product of orthogonal
Gaussians of equal width (on resonance).  As coupling 
increases, the wavepacket becomes stretched in a direction determined
by the angle $\gamma^{(1)}$, which on resonance is simply equal to 
$\pi/4$. This stretching increases up to $\lambda_c$, where the wavefunction
diverges.  We thus see the significance of the length $l_-$ 
introduced earlier - it is the extent of the wavefunction in the 
direction of this stretching.  Correspondingly, $l_+$ is the
extent of the wavefunction in the orthogonal direction.

In the super-radiant phase, 
the ground state is degenerate.  We shall initially consider
the ground-state wavefunction of
the effective Hamiltonian $H^{(2)}$ with
displacements chosen in Eq. (\ref{disp1}).
This is diagonal in the  $Q_{1}$-$Q_{2}$ representation of
Eq. (\ref{XYQ1Q2}), and therefore its ground-state wavefunction is
\begin{eqnarray}
\Psi^{(2)}_{G}\rb{Q_{1},Q_{2}}
= G_-^{(2)}\rb{Q_{1}}G_+^{(2)}\rb{Q_{2}}.
\end{eqnarray}
Using Eqns. (\ref{x-ycoord}),(\ref{disp1}), (\ref{XY}),
and (\ref{XYQ1Q2}) we may write this in 
the original $x$-$y$ representation as
\begin{eqnarray}
\Psi^{(2)}_{G}\rb{x,y} &=& 
  G_-^{(2)}\rb{  \rb{x -\sqrt{2\alpha/\omega}}\cos\gamma^{(2)} 
               - \sqrt{\omega_0/\tilde{\omega}}
	         \rb{y + \sqrt{2 \beta /\omega_0}}\sin\gamma^{(2)}} 
 \nonumber \\
&&\times 
  G_+^{(2)}\rb{  \rb{x - \sqrt{2\alpha/\omega}}\sin\gamma^{(2)} 
               + \sqrt{\omega_0/\tilde{\omega}}
		 \rb{y + \sqrt{2 \beta /\omega_0}}\cos\gamma^{(2)}}.
\end{eqnarray}
This expression contains displacements involving the macroscopic 
quantities $\alpha$ and $\beta$, and so we
define the new coordinates $x'$ and $y'$ to remove them:
\begin{eqnarray}
x' \equiv  x - \Delta_x;~~
y' \equiv y + \Delta_y ,
\label{xyprime1}
\end{eqnarray}
with
\begin{eqnarray}
\Delta_x \equiv \sqrt{2\alpha/\omega};~~
\Delta_y \equiv \sqrt{2 \beta /\omega_0},
\end{eqnarray}
which are both proportional to $\sqrt{j}$.
The relationship between the coordinate system $X$-$Y$ of Eq. (\ref{XY}) 
and $x'$-$y'$ is 
very simple, namely $x'=X$ and $y'=\sqrt{\tilde{\omega}/\omega_0}~Y$.
The coordinate system $x'$-$y'$ is useful because 
although $X$-$Y$ is the diagonal
representation for the super-radiant phase, the definition of 
these coordinates depends upon $\tilde{\omega}$ and hence upon 
$\lambda$, which distorts the picture.
In terms of these coordinates the wavefunction becomes
\begin{eqnarray}
\Psi^{(2)}_{G}\rb{x',y'} &=&
\rb{\omega_0 / \tilde{\omega}}^{1/4}
  G_-^{(2)}\rb{x'\cos\gamma^{(2)}
	       -\sqrt{\omega_0/\tilde{\omega}} y'\sin\gamma^{(2)}} 
  \nonumber \\
&&\times 
  G_+^{(2)}\rb{x'\sin\gamma^{(2)} 
               + \sqrt{\omega_0/\tilde{\omega}}y'\cos\gamma^{(2)}}.
\end{eqnarray}
%%%%%%%%%%%%%%%%%%%%%%%%%%%%%%%%%%%%%%%%%%%%%%%%%%%%%%%%%%%%%%%%%%
\begin{figure}[tb]
  \centerline{
  \includegraphics[clip=true,width=0.8\textwidth]{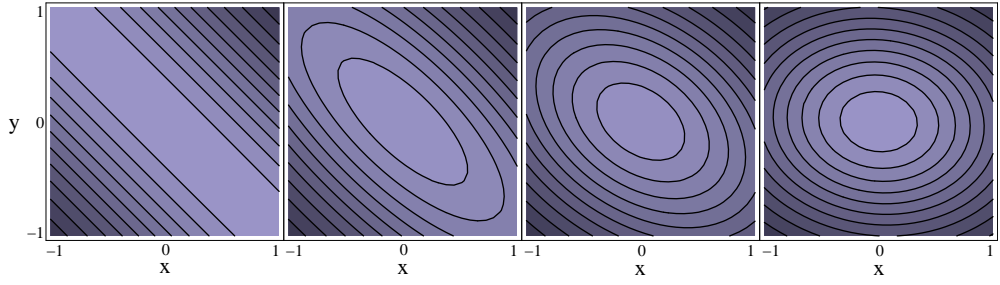}}
   \caption{\label{HCwavefn}
    The ground-state wavefunction $\Psi_G^{(2)}$ of the 
    high-coupling Hamiltonian $H^{(2)}$
    in the $x'$-$y'$ position-momentum representation 
    for couplings $\lambda = 0.500001,0.51,0.6,1.0$.
    The Hamiltonian is resonant: $\omega = \omega_0 = 1$, 
    $\lambda_c=0.5$. 
 }
\end{figure}
%%%%%%%%%%%%%%%%%%%%%%%%%%%%%%%%%%%%%%%%%%%%%%%%%%%%%%%%%%%%
Figure \ref{HCwavefn} shows $\Psi_{G}^{(2)}\rb{x',y'}$ for four different
couplings. Just above the phase transition the wavefunction is in a 
highly deformed state, characterised by the divergent $l_-$.
As the coupling increases further above $\lambda_c$, the wavefunction 
relaxes back to a well localised state.

\begin{figure}[bt]
\centerline{\includegraphics[clip=true,width=0.5\textwidth]{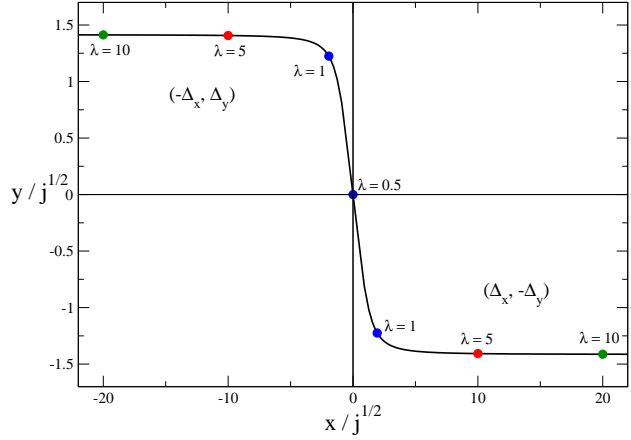}}
\caption{\label{dxdy} Parametric plot of the (scaled) displacements
$\rb{\Delta_x,-\Delta_y}$ and $-\rb{\Delta_x,\Delta_y}$
as $\lambda$ is
varied between 0.5 and 10.
The Hamiltonian is resonant; $\omega = \omega_0 = 1$.  }
\end{figure}

When considered in the original $x$-$y$ representation, the
wavefunction $\Psi_{G}^{(2)}\rb{x',y'}$ pictured in Fig. \ref{HCwavefn}
is centered about the 
point $\rb{+\Delta_x,-\Delta_y}$, which lies in the lower-right
quadrant of the $x$-$y$ plane.  The complementary wavefunction, 
identical in shape with this one but determined by the displacements 
(\ref{disp2}), is centered at 
$\rb{-\Delta_x,+\Delta_y}$, in the upper-left quadrant.
The positions of these two centers as parametric functions of coupling
are shown in Fig. \ref{dxdy}.  These two 
wavefunctions, corresponding to the two choices of displacement, 
are separated from the 
origin by an amount proportional to $\sqrt{j}$.
%%%%%%%%%%%%%%%%%%%%%%%%%%%%%%%%%%%%%%%%%%%%%%%%%%%%%%%%%%%%
%%%%%%%%%%%%%%%%%%%%%%%%%%%%%%%%%%%%%%%%%%%%%%%%%%%%%%%%%%%%
It is thus clear that neither of these wavefunctions is symmetric 
under rotation of $\pi$ about the origin of the $x$-$y$ coordinate system,
demonstrating once more that the $\Pi$ symmetry has been broken.  There is,
however, symmetry with respect to a rotation of $\pi$ about the origin of 
each $x'-y'$ coordinate system, which corresponds to
the existence of the local 
symmetries associated with $\Pi^{(2)}$ of  Eq. (\ref{Pi2}).

It is interesting to consider the behaviour of the 
ground-state wavefunction as $\lambda \go \infty$. 
In this limit $\varepsilon^{(2)}_- \go \omega$,
$\varepsilon^{(2)}_+ \go 4\lambda^2/\omega$
and the mixing angle of the two modes $\gamma^{(2)}$ tends to zero,
meaning that the modes decouple.  The Bogoliubov 
transformations of the modes become
\begin{eqnarray}
e_{1}^\dag \go c^\dag~~~&;&~~~e_{1} \go c\nonumber\\
e_{2}^\dag \go \frac{1}{2\sqrt{2}}\rb{3 d^\dag + d}~~~&;&~~~
e_{2}\go \frac{1}{2\sqrt{2}}\rb{d^\dag + 3d},
\label{laminf}
\end{eqnarray}
illustrating the decoupling.  Note that the
$e_{1}$ simply reverts to the $c$ mode, whereas the 
$e_{2}$ mode tends towards a linear combination of the 
annihilation and creation operators.  
In this limit, the wavefunction becomes
\begin{eqnarray}
  \Psi^{(2)}_{G}\rb{x',y'} &\go& 
  (\frac{\omega \omega_0}{2 \lambda^2})^{1/4}
  G^{(2)}_-\rb{x'}G^{(2)}_+\rb{\frac{\sqrt{2}\lambda_c}{\lambda} y'}
  \nonumber \\
  &=& \sqrt{\frac{2\lambda_c}{\pi}} 
  \exp\rb{-\omega_0 y'^2 - \frac{\omega}{2}x'^2},
\end{eqnarray}
which is independent of $\lambda$.

\subsection{Squeezing \label{sqg}}
 
A bosonic field may said to be squeezed if the uncertainty in
either of its quadratures, ($x$ or $p_x$) 
is less than the uncertainty in a coherent state \cite{ro:gl}.  
A coherent state
is a minimum uncertainty state with
$\rb{\Delta x}^2 \rb{\Delta p_x}^2 = 1/4$ and 
with the uncertainty apportioned evenly between 
the two quadratures.  Therefore,
the field is squeezed whenever $\rb{\Delta x}^2$ or $\rb{\Delta p_x}^2$ 
has a value lower than  $1/2$ \cite{wa:mi}.

We define the two quadrature variances of the original field mode $a$ 
by $\rb{\Delta x }^2 \equiv \ew{x^2}-\ew{x}^2$ and 
$\rb{\Delta p_x }^2 \equiv \ew{p_x^2}-\ew{p_x}^2$, which may be 
seen to be equal to
\begin{eqnarray}
\rb{\Delta x }^2 = \frac{1}{2\omega} 
  \left\{ 
    1+\ew{{a^\dag}^2}+\ew{a^2}+2\ew{a^\dag a}+\rb{\ew{a^\dag}+ \ew{a}}^2
  \right\},
\nonumber \\
\rb{\Delta p_x }^2 = \frac{\omega}{2} 
  \left\{ 
    1-\ew{{a^\dag}^2}-\ew{a^2}+2\ew{a^\dag a}+\rb{\ew{a^\dag}+ \ew{a}}^2
  \right\}.\label{xvar}
\end{eqnarray}
As we have introduced a bosonic algebra for the atomic collection,
we now introduce an analogous definition for squeezing in the atoms, and say 
that in terms of the variances,
\begin{eqnarray}
\rb{\Delta y }^2 = \frac{1}{2\omega_0} 
  \left\{ 
    1+\ew{{b^\dag}^2}+\ew{b^2}+2\ew{b^\dag b}+\rb{\ew{b^\dag}+ \ew{b}}^2
  \right\},
\nonumber \\
\rb{\Delta p_y }^2 = \frac{\omega_0}{2} 
  \left\{ 
    1-\ew{{b^\dag}^2}-\ew{b^2}+2\ew{b^\dag b}+\rb{\ew{b^\dag}+ \ew{b}}^2
  \right\}.\label{yvar}
\end{eqnarray}
the atoms are squeezed if either $\rb{\Delta y }^2$ or 
$\rb{\Delta p_y }^2$ is less than $1/2$.
The squeezing of atomic ensembles is usually defined in terms 
of the collective operators \cite{wazo:sqg,ueda:sqg}. Because the
angular momentum operators obey the commutation relation, 
$\left[J_+,J_-\right]=2J_z$, the uncertainty 
relation,
$\rb{\Delta J_x}^2 \rb{\Delta J_y}^2 \ge \frac{1}{4}|\ew{J_z}|^2$
holds for any state.
By substituting in the Holstein-Primakoff forms
into this expression
and taking the thermodynamic limit, we see that this relation 
reduces to $\rb{\Delta y}^2 \rb{\Delta p_y}^2 \ge 1/4$, 
demonstrating the equivalence in the thermodynamic limit
of our definition in terms of $y$ and $p_y$ and the usual one.

%%%%%%%%%%%%%%%%%%%%%%%%%%%%%%%%%%%%%%%%%%%%%%%%%%%%%%%%%%%%
\begin{figure}[bt]
  \centerline{\includegraphics[clip=true,width=0.6\textwidth]
    {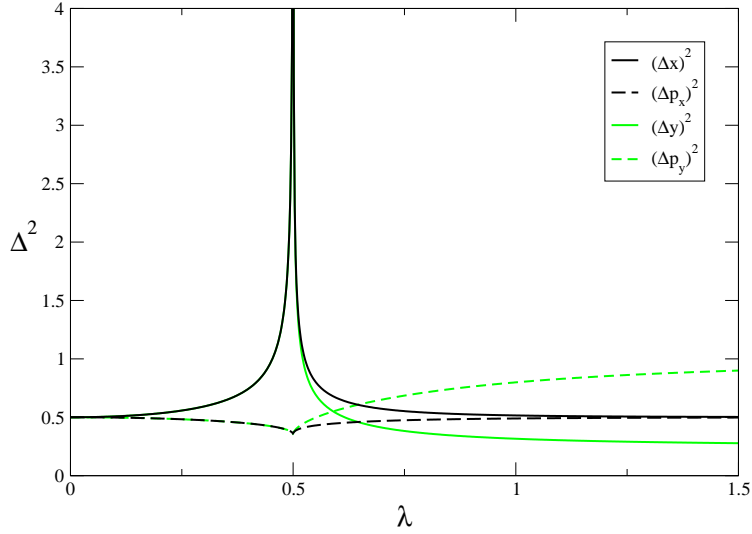}}
  \caption{\label{sqzg}
    The squeezing variances of the ground state of the 
    DH in the thermodynamic limit. 
    The Hamiltonian is resonant: $\omega = \omega_0 = 1$, 
    $\lambda_c=0.5$. Note that on resonance, $\rb{\Delta x}^2$ and 
    $\rb{\Delta y}^2$ are coincident for $\lambda<\lambda_c$, and the same
    for $\rb{\Delta p_x}^2$ and $\rb{\Delta p_y}^2$
  }
\end{figure}
%%%%%%%%%%%%%%%%%%%%%%%%%%%%%%%%%%%%%%%%%%%%%%%%%%%%%%%%%%%%
In the normal phase the expressions for the variances are evaluated by 
simply making the appropriate substitutions from Appendix \ref{appBOGT} 
and taking their ground-state expectation value.
In the super-radiant phase, it can be shown that the variances of the
original field and atomic modes of Eqs. (\ref{xvar}) and (\ref{yvar}) can 
be expressed in terms of the displaced coordinates as follows
\begin{eqnarray}
  \rb{\Delta x }^2   =  \rb{\Delta x' }^2   =  \rb{\Delta X }^2
  &;&~~~~~~~~
  \rb{\Delta p_x }^2 =  \rb{\Delta p_x' }^2 =  \rb{\Delta P_X }^2,
  \nonumber \\
  \rb{\Delta y }^2   =  \rb{\Delta y' }^2    
                     =  \sqrt{\tilde{\omega}/\omega_0}\rb{\Delta Y }^2
  &;&~~~~~~~~~
  \rb{\Delta p_y }^2 =  \rb{\Delta p_y' }^2  
                     =  \sqrt{\tilde{\omega}/\omega_0}\rb{\Delta P_Y }^2.
\end{eqnarray}
This results from the fact that the squeezing variances do
not depend upon the displacements of the field modes, and thus
evaluating the super-radiant variances is as simple as in the normal 
phase.

The analytic values of these variances in the ground state 
are shown in Appendix \ref{sqzgvar} and are
plotted as functions of coupling in Fig. \ref{sqzg}.
In the normal phase, as $\lambda$ approaches
$\lambda_c$ there is a sharp increase, and eventually a divergence, in 
$\rb{\Delta x }^2$ and $\rb{\Delta y }^2$.  This is 
accompanied by a slight squeezing of the momentum variances.  In the 
super-radiant phase the initially divergent values of 
$\rb{\Delta x }^2$ and $\rb{\Delta y }^2$ reduce rapidly 
with increasing coupling.  
The behaviour of these variances reflects the nature of the
wavefunctions plotted in Figs. \ref{LCwavefn} and \ref{HCwavefn}.
Notice that as $\lambda \go \infty$, $\rb{\Delta x }^2$ and 
$\rb{\Delta p_x }^2$ return to their $\lambda=0$ values, whereas  
$\rb{\Delta y }^2$ and $\rb{\Delta p_y }^2$ become squeezed and 
anti-squeezed respectively.  
This is in agreement with the results of Eq. (\ref{laminf}), which 
show that the $e_1$ mode becomes identical to the $c$-mode, which is 
unsqueezed, whereas the $e_2$ mode reverts to a linear superposition of
$d$ and $d^\dag$ operators, which is a specific example of the Bogoliubov
transformation producing a squeezed state \cite{bi:v2}.

% LocalWords:  DH Eq QPT Hamiltonians Primakoff bosonic Dicke bi diagonalised
% LocalWords:  wavefunctions diagonalise quantise eigenstates diagonalisation
% LocalWords:  co ordinates requantisation decoupled Bogoliubov transfomations
% LocalWords:  dislpacements eigenfunctions behaviour photonic towards bose xy
% LocalWords:  wavefunction Gaussians wavepacket normalised Parametric decouple
% LocalWords:  decoupling quadratures superposition polariton characterised
% LocalWords:  localised parametric unsqueezed

\newpage
\section{The Onset of Chaos \label{Ochaos}}
  As we have just demonstrated, the DH is exactly integrable in the 
thermodynamic limit.  However, for finite $j$, this is not the case 
and the possibility of quantum chaos remains.
The signature of quantum chaos that we use to
investigate this possibility
is the character 
of the energy spectrum as quantified by the nearest-neighbour level 
distribution $P\rb{S}$.  
Bohigas et al.\cite{bo:gs} first conjectured that the study 
of spectral quantities such as $P\rb{S}$, and their comparison
with the results from random matrix theory should give an 
indicator of quantum chaos.  This  may 
be understood by the following argument.
Classically integrable systems have high degrees of symmetry and hence 
their quantum counterparts have
many conserved quantum numbers.  This permits level-crossings 
to occur in the spectrum, leading to a $P\rb{S}$ with a maximum
at small level-spacing, $S \rightarrow 0$, with a $P\rb{S}$ given by 
the Poissonian distribution $P_\mathrm{P}\rb{S} = \exp\rb{-S}$.
We shall call quantum spectra with Poissonian statistics ``quasi-integrable''.
Conversely, classically chaotic systems have no 
such integrals of motion and we thus expect their quantum energy spectra to 
be highly correlated and 
absent of crossings, leading to $P\rb{S}\rightarrow 0$ 
as $S\rightarrow 0$.  Although the precise form of the $P\rb{S}$ 
for chaotic systems depends on 
the symmetries of the model, we shall find that only the
Wigner-Dyson distribution, 
$P_\mathrm{W}\rb{S} = \pi S/2 \exp\rb{-\pi S^2/4}$,
is of relevance here \cite{guhr}.

Despite its popularity, it should be pointed out that the 
correspondence between the $P \rb{S}$ distribution and the integrability 
or otherwise of the classical system is not absolute, and 
exceptions do exist \cite{ccg:except,wvfs:except}. Despite this,
the $P\rb{S}$ does provide a convenient and useful
indicator of quantum chaos, and the conjecture does hold true
in a countless examples.
In the case in hand,
this signature turns out to be very accurate, 
as will be evinced when we compare the $P\rb{S}$ results with those
of our semi-classical model.

\subsection{Numerical diagonalisation}

Exact solutions for the DH at finite $j$ do not exist, except 
in the very special case of $j=1/2$ where isolated exact 
(``Juddian'') solutions may be found \cite{ju:dd,ce:jd}.  Consequently 
we employ numerical
diagonalisation to investigate the system.
To perform these diagonalisations we use the basis
$\left\{\ket{n}\otimes \ket{j,m}\right\}$, 
where $\ket{n}$ are number states of the field, 
and $\ket{j,m}$ are 
the Dicke states.  In performing 
the diagonalisation, we truncate the bosonic Hilbert space
but always maintain the full Hilbert space of
the pseudo-spin.  
The size of the matrices requiring diagonalisation is reduced by
restricting ourselves to a single parity subspace, which is 
achieved by only considering states 
with $n + m + j$ even or odd for positive and negative parity respectively.
With $j$ finite, $\Pi$ is a good quantum number, 
independent of coupling, and the ground state
always has positive parity.

The results obtained via this diagonalisation 
for the ground-state energy and its second derivative
for a sequence of finite $j$ values have been 
plotted alongside the $j \go \infty$ results in Fig. \ref{gse}, whilst 
the corresponding
atomic inversions and mean photon numbers are plotted
in Fig. \ref{aipno}.  These figures demonstrate how rapidly the finite
$j$ results approach their thermodynamic limits as $j$ is
increased.

\subsection{Level statistics}
Having numerically obtained energy spectra of the DH, we can construct
the nearest-neighbour level-spacing distribution $P\rb{S}$.  This is
formed from a large number of levels from the
spectrum, which we initially unfold to rid of 
secular variation \cite{guhr}.  We then
calculate the level spacings,
\begin{eqnarray}
  S_n = E_{n+1} - E_n, 
\end{eqnarray}
where $\left\{E_n;~n=0,1,\ldots\right\}$ is the set of 
eigen-energies of the DH with positive parity, 
and construct their distribution function $P\rb{S}$.   Finally, we 
normalise the results for comparison with the universal ensembles 
of Random Matrix Theory \cite{guhr}.

%%%%%%%%%%%%%%%%%%%%%%%%%%%%%%%%%%%%%%%%%%%%%%%%%%%%%%%%%%%%%%%%%%%%%%%%%%%%%
\begin{figure}[tb]
  \centerline{\includegraphics[clip=true,width=0.6\columnwidth]
    {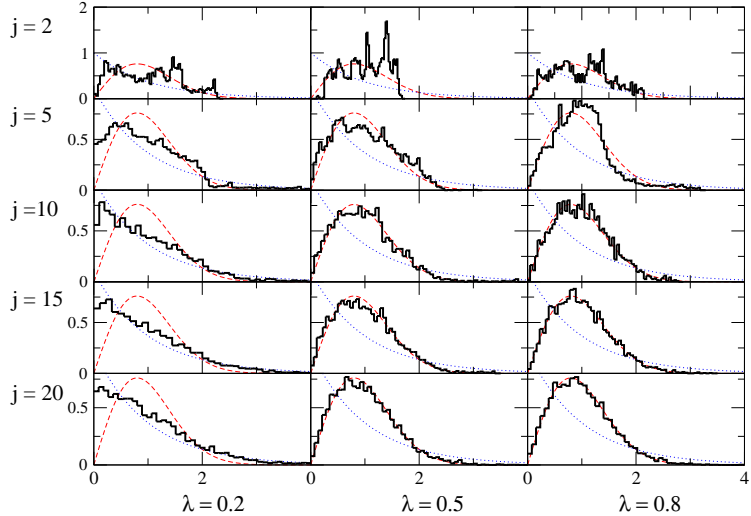}}
  \caption{\label{P(S)}
    Plots of nearest-neighbour distributions  
    $P(S)$ for the 
    Dicke Hamiltonian, for different couplings $\lambda$ and pseudo-spin $j$.
    Also plotted are the universal Poissonian (dots) to Wigner (dashes) 
    distributions.   
    The Hamiltonian is resonant: $\omega = \omega_0 = 1$, 
    $\lambda_c=0.5$. }
\end{figure}
%%%%%%%%%%%%%%%%%%%%%%%%%%%%%%%%%%%%%%%%%%%%%%%%%%%%%%%%%%%%%%%%%%%%%%%%%%%%

Figure \ref{P(S)} shows the $P\rb{S}$ distributions obtained for the
DH at various values of $\lambda$ and $j$.
At low $j$ ($j \le 3$) the $P\rb{S}$ clearly do not correspond to 
any of the universal ensembles, but rather to non-generic 
distributions consisting of several isolated peaks.  
This is most obvious in the $j=1/2$ case (not shown here), which is 
known as the Rabi Hamiltonian (RH) \cite{ii:ra}.  The RH
has a spectrum that is of ``picket-fence'' character 
\cite{kus}, which is characteristic of genuinely integrable models
such as one-dimensional systems and systems of harmonic oscillators
\cite{be:ta}.  
The RH is unusual and must be treated as a special case because, although
it has never been shown to be integrable, isolated exact solutions 
do exist \cite{ju:dd,ce:jd}. Moreover the model 
is separable and may be reduced to a single degree of freedom \cite{gr:ho1}.

Returning to the $P\rb{S}$ distributions, we see that at 
low couplings $\lambda < \lambda_c$, 
(for example, $\lambda=0.2$ in Fig. (\ref{P(S)})), 
as we increase $j$, the $P\rb{S}$ loses its non-generic 
features and 
approaches ever closer the Poissonian distribution, 
$P_\mathrm{P}\rb{S}$.
At and above the critical coupling 
($\lambda = 0.5$ and $0.8$ in Fig. \ref{P(S)})
the spectrum is seen to converge onto the Wigner-Dyson 
distribution $P_\mathrm{W}\rb{S}$ as $j$ is increased.  

\begin{figure}[tb]
  \centerline{
    \includegraphics[clip=true,width=0.6\columnwidth]{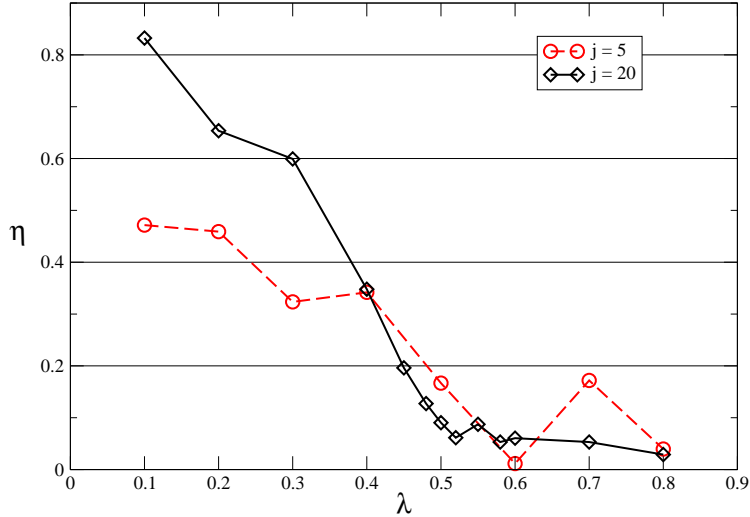}
   }
  \caption{\label{Pj20}  
    The modulus of $\eta$, Eq. (\ref{eta}), plotted as a function of 
    coupling for systems of $j=5$ and $j=20$. A value of $\eta =1$ 
    indicates Poissonian statistics and $\eta =0$ corresponds to 
    Wigner-Dyson.
    The system is on 
    scaled resonance ($\omega = \omega_0 =1$). }
\end{figure}

The nature of the change in the $P\rb{S}$ distribution
may be characterised by the quantity
\begin{eqnarray}
\eta \equiv \left|\frac{\int_0^{S_0} \left[ P\rb{S} - P_W\rb{S}\right] dS}
            {\int_0^{S_0}  \left[ P_P\rb{S} - P_W\rb{S} \right]dS}\right|,
\label{eta}
\end{eqnarray}
where $S_0 = 0.472913\ldots$, the value of $S$ at which the two 
generic distributions $P_P\rb{S}$ and  $P_W\rb{S}$ first intersect 
\cite{gesh:shards}.  $\eta$ measures the degree of similarity of the 
calculated 
$P\rb{S}$ to the Wigner surmise $P_W\rb{S}$, and is normalised such 
that if $P\rb{S} = P_W\rb{S}$ then $\eta =0$, and if
$P\rb{S} = P_P\rb{S}$ then $\eta =1$.  The behaviour of $\eta$ as a 
function of
coupling for $j=5$ and $j=20$ is shown in Fig. \ref{Pj20}.
Considering the $j=20$ case first we see that
the spectrum is strongly Poissonian at low couplings and that 
at $\lambda$ is increased towards $\lambda_c$, it becomes more 
Wigner-Dyson like.  This proceeds until we reach $\lambda_c$, 
about which
the spectrum is remarkably well described by $P_\mathrm{W}\rb{S}$.
Note that for $\lambda < \lambda_c$ the value of $\eta$ drops 
steadily with coupling, whereas above $\lambda_c$ it
maintains an 
approximately constant value close to zero.
For the $j=5$ case, a similar transition is observed, but  
it is not as pronounced and the agreement with the universal distributions
is not as good as in the higher $j$ case.

Thus, for sufficiently high $j$, we see a significant change in
$P\rb{S}$ as $\lambda$ is increased from zero through the critical value
$\lambda_c$.  Below $\lambda_c$, there is a significant amount of 
level-crossing, which decreases as we approach $\lambda_c$.  Above 
$\lambda_c$ there is practically none, to within statistical error. 
Thus we conclude that the
precursors of the QPT in this model lead to a cross-over from 
quasi-integrable to quantum chaotic behaviour 
at $\lambda \approx \lambda_c$ for sufficiently high $j$.

%%%%%%%%%%%%%%%%%%%%%%%%%%%%%%%%%%%%%%%%%%%%%%%%%%%%%%%%%%%%%%%%%%
%%%%%%%%%%%%%%%%%%%%%%%%%%%%%%%%%%%%%%%%%%%%%%%%%%%%%%%%%%%%%%%%%%%

\subsection{Regularity at low energy}

A further transition between integrable and chaotic
behaviour is observed in the sequence of level spacings $S_n$
as the coupling is increased from $\lambda_c$ to $\infty$.
In the $\lambda \rightarrow \infty$ limit, the DH is integrable for arbitrary
$j$ and equivalent to
\begin{eqnarray}
  H_{\lambda \go \infty} = \omega a^\dagger a 
     + 2\frac{\lambda}{\sqrt{2j}} \rb{a^\dagger + a}J_x.
\end{eqnarray}
The eigenstates of $H_{\lambda \go \infty}$ 
are obviously eigenstates of $J_x$, and thus
\begin{eqnarray}
  H_{\lambda \go \infty} = \omega a^\dagger a 
     + 2m\frac{\lambda}{\sqrt{2j}} \rb{a^\dagger + a},
\end{eqnarray}
where $m=-j,\ldots, j$ is the eigenvalue of $J_x$.  This bosonic Hamiltonian 
is diagonalised by the displacement $a \go a - 2m\lambda/\sqrt{2j}$,
giving the eigenvalues to be
\begin{eqnarray}
E_{km} = \frac{\omega}{j} k - \frac{2 \lambda^2}{\omega j^2} m^2,
\end{eqnarray}
where $k=0,1,2,\ldots$.  The energy levels with $+m$ and $-m$ are  degenerate.

\begin{figure}[tb]
  \centerline{
    \includegraphics[clip=true,width=0.6\columnwidth]
      {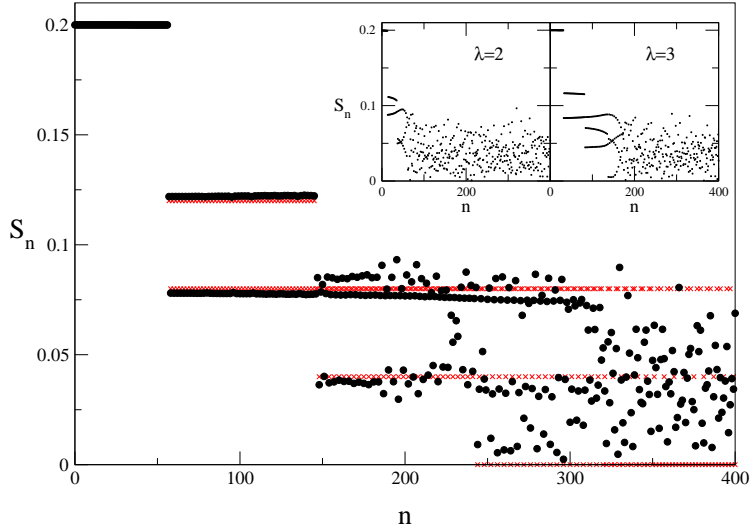}}
  \caption{\label{anom}
    Nearest-neighbour spacing 
    $S_n=E_{n+1}-E_n$ vs. eigenvalue 
    number $n$ plot for $j=5$ with $\lambda =4$.
    Horizontal crosses: results for the integrable 
    $\lambda \rightarrow \infty$ Hamiltonian.  Inset: $j=5$ results
    with $\lambda =2$ and $\lambda=3$.
    The Hamiltonian is resonant: $\omega=\omega_0=1$, $\lambda_c=0.5$
  }
\end{figure}

As $\lambda$ is increased from $\lambda_c$ to approach this 
$\lambda \go \infty$ limit
with $j$ fixed, the spectrum reverts from Wigner-like to integrable.  
However, it does not follow the usual transition sequence
of Wigner distribution gradually changing into a Poissonian one, 
as one might expect, 
but 
rather through a sequence illustrated by Fig. \ref{anom}.
For couplings sufficiently higher than $\lambda_c$, the spectrum 
becomes very regular at low energy, where it approximates the 
$\lambda \rightarrow \infty$ of results very closely. 
Outside the regular 
region the spectrum is well described by the Wigner surmise, and 
the energy-scale over which the change between the two regimes occurs 
is seen to be surprisingly narrow.  As coupling is increased, the 
size of the low-energy integrable window increases, until it 
eventually engulfs the whole spectrum as $\lambda \rightarrow \infty$.
This division of the spectrum into regions is close to Percival's 
conception of how regular and irregular behaviour would manifest
itself in quantum systems \cite{perc:1}.
%%%%%%%%%%%%%%%%%%%%%%%%%%%%%%%%%%%%%%%%%%%%%%%%%%%%%%%%%%%%%%%%%%%%%%%%%%%
%%%%%%%%%%%%%%%%%%%%%%%%%%%%%%%%%%%%%%%%%%%%%%%%%%%%%%%%%%%%%%%%%%%%%%%%%%%%%

\subsection{Wavefunctions for finite $j$}

We now consider the wavefunctions of the DH at finite $j$.  To 
do this, we shall use the position-momentum representation 
of Eq. (\ref{x-ycoord})
used earlier in discussing the wavefunctions in the 
thermodynamic limit.  We begin with the eigenfunctions 
obtained from numerical
diagonalisation, which are of the form
\begin{eqnarray}
  \ket{\Psi_{nm}} = \sum_{n=0}^{n_c}\sum_{m=-j}^{+j} 
    c^{(j)}_{nm}\ket{n}\ket{j,m},
\end{eqnarray}
where $n_c$ is the maximum boson number in the artificially 
truncated Fock space, and $c^{(j)}_{nm}$ are coefficients.  The position
representatives of the number states of the field $\ket{n}$ are simply 
the usual Harmonic oscillator eigenfunctions 
\begin{eqnarray}
  \ew{x|n}= 
    \frac{1}{2^n~n!}
    \sqrt{\frac{\omega}{\pi}}
    e^{-\frac{1}{2}\omega x^2}
    H_n\rb{\sqrt{\omega} x}
\label{oei}
\end{eqnarray}
where $H_n$ is the $n$th Hermite polynomial.  For 
the angular momentum part of the basis vector, we recall that 
under the Holstein-Primakoff mapping $J_z \go b^\dag b - j$, 
and thus the Dicke states are eigenstates of 
$b^\dag b$ with eigenvalue $\rb{j+m}$: 
$b^\dag b \ket{j,m} = \rb{j+m}\ket{j,m}$,
$-j\le m \le j$.  Consequently, we may represent the Dicke states 
in the same way as the Fock states above,
allowing us to write 
the total wavefunction in the two-dimensional 
position representation as
\begin{eqnarray}
  \Psi_{nm}\rb{x,y} =
%  = \ew{x,y|\Psi_{nm}} 
    \frac{\sqrt{\omega\omega_0}}{\pi}
    e^{-\frac{1}{2}\rb{\omega x^2 + \omega_0 y^2}}
   \sum_{n=0}^{n_c}\sum_{m=-j}^{+j} c^{(j)}_{nm}
    \frac{ H_n\rb{\sqrt{\omega} x}
    H_{j+m}\rb{\sqrt{\omega_0} y}}{2^{\rb{n+j+m}}~n!\rb{j+m}!}.
\end{eqnarray}
This is a very productive representation
in which to study this Hamiltonian.  It does however, suffer from the 
drawback that whereas 
the set of oscillator eigenfunctions Eq. (\ref{oei}) forms
an orthonormal set in the $x$ direction, this is not the case in the 
$y$ direction as we only keep up to the ($2j$)th oscillator eigenfunction
in this direction.  This means, for example, that we could not go 
from an arbitrary wavefunction in the $y$ direction to a description
in terms of the Dicke states, because we do not have a complete set of 
functions in this direction.
Specifically, the significant width of the wavefunction is limited 
in the $y$ direction by the maximum significant extent  of 
the highest Hermite polynomial $H_{2j}$.
However, if we know the value of $j$ and only consider wavefunctions that
are describable in terms of these then the representation is unique.
%%%%%%%%%%%%%%%%%%%%%%%%%%%%%%%%%%%%%%%%%%%%%%%%%%%%%%%%%%%%%%%%%%%%%%%%%
\begin{figure}[tb]
  \centerline{
    \includegraphics[clip=true,width=0.4\textwidth]{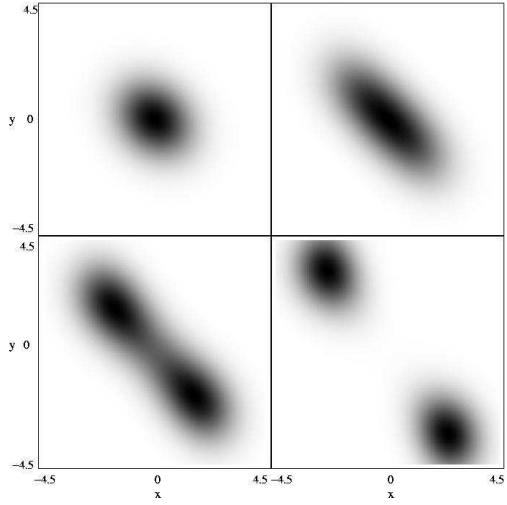}}
  \caption{
  \label{wavefn}
     The modulus of the ground-state wavefunction 
    $\psi\rb{x,y}$
    of the Dicke Hamiltonian in the abstract $x$-$y$ representation  
    for finite $j=5$, at couplings of 
    $\lambda / \lambda_c$ = 0.2, 0.5, 0.6, 0.7.  Black corresponds 
    to $\mathrm{Max}|\psi|$ and white corresponds to zero.  The 
    Hamiltonian is resonant $\omega =\omega_0=1$; $\lambda_c=0.5$.
  }
\end{figure}
%%%%%%%%%%%%%%%%%%%%%%%%%%%%%%%%%%%%%%%%%%%%%%%%%%%%%%%%%%%%%%%%%%%%%%%%%

Figure \ref{wavefn} shows the ground-state 
wavefunction of DH with $j=5$, for a series of increasing couplings.  
Note that the wavefunction is 
always invariant under a rotation of $\pi$ about the 
origin as demanded by the $\Pi$ symmetry.
This wavefunction starts as a single lobe centred at 
the origin for low coupling.  As the coupling increases,
the two modes start mixing, leading to a stretching of the 
single-peaked wavefunction, which then splits into two as coupling is 
increased through a coupling approximately equal to $\lambda_c$.  
With further increases in coupling the two lobes 
move away from each other in their respective quadrants of the $x-y$ plane.

The key observation regarding the two lobes formed above $\lambda_c$ is that,
provided $j$ and $\lambda$ are both sufficiently large, their 
displacement from each other 
is proportional to $\sqrt{j}$, and that this is 
a macroscopic quantity.  The excited states exhibit similar behaviour, 
having an extent proportional to $\sqrt{j}$ above the phase transition.

Therefore, around the critical coupling 
$\lambda \approx \lambda_c$, the wavefunctions of the 
system become delocalised, and the extent of this 
delocalisation is proportional to $\sqrt{j}$.  As this is a macroscopic
quantity, we see that above $\lambda_c$, the system at finite $j$ develops
macroscopic coherence in its wavefunctions. The most
striking example of this is the ground state, where the two 
macroscopically different lobes are reminiscent of the two 
states of a Schr\"odinger's cat.

The delocalisation and accompanying macroscopic coherence are 
rather general features of the onset of chaos, and are
natural consequences of the 
exponential divergence of trajectories 
in a classically chaotic system \cite{zurek:1}.  
If we consider a small 
volume of initial conditions in the classical phase space 
(a well-localised initial wave-packet), 
and let the system evolve chaotically, 
this initial volume rapidly becomes  smeared out over the entire 
phase-space accessible to it. This is reflected in the quantum system by 
the delocalisation of the wavefunctions.
That such systems are macroscopic coherent may be seen from 
the observation that under Hamiltonian dynamics, the volume of the 
initial ``wave-packet'' remains constant in time (Liouville's theorem).  
This means that the exponential divergence in some direction leads 
to the exponential contraction in others.  This contraction will continue
until the size of the packet becomes of the order of $\hbar$ and 
quantum effects come into play.  If we imagine that the 
wave-packet becomes narrow in the direction of momentum $p$, then 
the uncertainty $\Delta p$ becomes very small.  In order that the 
Heisenberg uncertainty relation holds, the uncertainty in the 
corresponding coordinate $\Delta x$ must become very large, and 
this leads to the emergence of macroscopic coherence in the system. 

This effect is what was observed in the variances 
calculated earlier in connection with squeezing 
in the thermodynamic limit.  As $\lambda \go \lambda_c$
from below, the variances $\rb{\Delta x}^2$ and $\rb{\Delta y}^2$   
diverged, with $\rb{\Delta p_x}^2$ 
and $\rb{\Delta p_y}^2$ 
remaining near their quantum limit of $1/2$.  The behaviour 
of these variances then reflects the onset of quantum chaos 
and the macroscopic coherence of the wavefunctions.
A vital difference between the $j \go \infty$ and the finite $j$ results
thus emerges in the super-radiant phase.  In the 
thermodynamic limit, the variances  $\rb{\Delta x}^2$ and 
$\rb{\Delta y}^2$ reduce as $\lambda$ is increased from $\lambda_c$, 
indicating that the wavefunctions become localised and lose this 
macroscopic coherence.  
Contrast this with the finite $j$, where sufficiently above $\lambda_c$ 
the wavefunction is always delocalised and the variances 
are ${\cal O} \rb{\sqrt{j}}$.  This is because, whereas at finite
$j$ we obey $\Pi$ symmetry and thus have both of 
the lobes of the wavefunction, in the
thermodynamic limit, we consider each lobe separately under 
the broken symmetry.  
The individual lobes are themselves localised and this is
where the discrepancy comes from.  This is, we believe,
the reason why although the spectrum is of the
Wigner-Dyson type for large $j$, the
spectrum in the $j\go \infty$ limit is integrable, 
as in this limit the wavefunctions possess no 
delocalisation and no macroscopic coherence.

This picture also provides us with an explanation of why
the $P\rb{S}$ for very small $j$ are of the non-generic
one-dimensional type.  As the extent of the wavefunction
in the $y$-direction is effectively constrained by the
number of harmonic eigenfunctions 
in that direction, which is determined by $j$, having a 
small $j$ prevents full delocalisation in this direction,
inhibiting the chaoticity of the quantum system.

% LocalWords:  diagonalisation DH diagonalisations Dicke bosonic neighbour Rabi
% LocalWords:  spacings normalise integrable Poissonian Dyson QPT behaviour Eq
% LocalWords:  characterised normalised towards eigen Wavefunctions boson Fock
% LocalWords:  wavefunctions eigenfunctions th Primakoff eigenstates Schr zurek
% LocalWords:  wavefunction orthonormal eigenfunction localisation odinger's nm
% LocalWords:  delocalisation macroscopically prange integrability Juddian et
% LocalWords:  localised wavepacket decoherence superpositions delocalised al
% LocalWords:  Bohigas dS diagonalised Liouville's chaoticity utilise

\newpage
\section{The Semi-Classical Model \label{clas}}
  
As noted in the introduction, there have been many different semi-classical
models derived from the DH \cite{fu:ru,mil,msln,fingea}. 
That there have been so 
many different approaches is a reflection of the fact that the 
quantum mechanical spin possesses no direct classical analogue.  
Nevertheless, semi-classical models can be constructed, and in the 
following we shall propose a new approach.  Before this, let 
us briefly examine some of the previous work.

A widely discussed approach is that of a Hartree-Fock 
type approximation in which one derives the Heisenberg equations of 
motion for the system and replaces the operators in these equations 
by their expectation values \cite{gr:ho1}.  
These are treated as classical variables and a set of non-linear 
equations of motion are obtained for them, 
which show classical chaos for certain 
parameter ranges \cite{mil}.  Despite this, the above approach is not
completely satisfactory as the motion only depends $j$ in a 
trivial way. Furuya at al. have studied a classical model 
similar to the one we propose below \cite{fu:ru}.  They derived their 
semi-classical
Hamiltonian by evaluating the expectation value of the DH
in a state composed of a product of photonic and atomic 
coherent states, and this system was also shown to exhibit chaos.  
Despite the similarity of their model to 
ours, they did not discuss the role of the phase transition in determining 
the chaoticity of the model, which is a key feature of our model.

We start with the DH in the bosonic form of Eq. (\ref{DHam2}):
\begin{eqnarray}
H = \omega_0 \rb{b^\dagger b - j} + \omega a^\dagger a
+ \lambda \rb{a^\dagger + a}
\rb{
  b^\dagger \sqrt{1-\frac{b^\dagger b}{2j}}  
  + \sqrt{1-\frac{b^\dagger b}{2j}}b }
\label{clDHam1}.
\end{eqnarray}
By using the inverse of the relations in Eq. (\ref{x-ycoord}), namely
\begin{eqnarray}
  a \equiv \sqrt{\frac{\omega}{2}}\rb{x + \frac{i}{\omega_0} p_x}
    &;&a^\dag \equiv  \sqrt{\frac{\omega}{2}}\rb{x - \frac{i}{\omega_0} p_x}
  \nonumber \\ 
  b \equiv \sqrt{\frac{\omega_0}{2}}\rb{y + \frac{i}{\omega_0} p_y}
    &;&b^\dag \equiv \sqrt{\frac{\omega_0}{2}}\rb{y - \frac{i}{\omega_0} p_y}
  ,
\end{eqnarray}
we may write this Hamiltonian in the position-momentum 
representation,
\begin{eqnarray}
H &=& -j \omega_0 
   + \frac{1}{2} 
   \rb{
        \omega^2 x^2 + p_x^2 - \omega +  \omega_0^2 y^2 + p_y^2 - \omega_0
      }
\nonumber \\
    &&+ \lambda \sqrt{\omega \omega_0} x 
        \left\{ 
		\rb{y - \frac{i}{\omega_0} p_y} \sqrt{1 -\eta} 
	        + \sqrt{1 -\eta} \rb{y + \frac{i}{\omega_0} p_y}
	\right\},
\end{eqnarray}
where we have written 
\begin{eqnarray}
  \eta = \rb{\omega_0^2 y^2 + p_y^2 - \omega_0} / \rb{ 4 j \omega_0}.
\end{eqnarray}
We now move very naturally from this
quantum-mechanical Hamiltonian to a semi-classical one by setting the 
position-momentum commutators to zero, i.e. 
$\left[x,p_x \right] =0$, $\left[y,p_y \right] =0$.  This causes 
the interaction term to become real and in terms of
classical variables we have
\begin{eqnarray}
H_\mathrm{sc}= -j \omega_0 
   &+& \frac{1}{2} 
   \rb{
        \omega^2 x^2 + p_x^2 - \omega +  \omega_0^2 y^2 + p_y^2 - \omega_0
      }
\nonumber \\
    &+& 2 \lambda \sqrt{
                        \omega \omega_0} 
                        ~x y \sqrt{1 -\frac{
                                          \omega_0^2 y^2 
                                           + p_y^2 - \omega_0}
                                           {4 j \omega_0}
                       }.
\label{Hcl}
\end{eqnarray}

Unusually, this Hamiltonian
contains an intrinsic constraint, which is determined by the requirement that
the square-root must remain real for the system to remain Hamiltonian.
This means that the inequality
\begin{eqnarray}
\eta = \frac{1}{4j \omega_0}\rb{\omega_0^2 y^2 + p_y^2 - \omega_0} \le 1
\label{etacond}
\end{eqnarray}
is satisfied for all time.

\subsection{Classical Phase Transition}

The Hamiltonian $H_\mathrm{sc}$ undergoes a spontaneous symmetry-breaking
phase transition that is directly analogous to the QPT of the quantum model.  
The exact correspondence between the classical 
and quantum Hamiltonians in the thermodynamic limit is because 
in this limit the system is exactly described with a mean-field
theory as used earlier, and the use of classical 
variables as we have done here is equivalent to a mean-field theory.
Consequently, we are able to derive classical effective Hamiltonians 
exactly as we did in the quantum case.  The
effective Hamiltonian for the normal phase is derived 
by simply letting $j \go \infty$ (i.e. $\eta \go 0$) in the 
Hamiltonian of Eq. (\ref{Hcl}).  This gives us
\begin{eqnarray}
H^{(1)}_\mathrm{sc} = \frac{1}{2}
\left\{ \omega^2 x^2 + p_x^2 + \omega_0^2 y^2 + p_y^2 
+ 4 \lambda \sqrt{\omega \omega_0}~ x y - \omega_0 - \omega\right\}
-j \omega_0,
\end{eqnarray}
which is identical to Eq. (\ref{H1jinf}) from the quantum analysis, 
and may be diagonalised with the same
rotation.
The equilibrium position of Hamiltonian $H^{(1)}_\mathrm{sc}$ is the origin:
$x=y=$ $p_x=p_y=0$.

An effective Hamiltonian for the super-radiant phase 
is derived in the same way as in the quantum case, by displacing the 
co-ordinates as in Eq. (\ref{xyprime1}), so that
$x \go x^\prime \pm \Delta_x$,$y \go y^\prime \mp \Delta_y$,
where the displacements are the same as before:
$\Delta_x \equiv \sqrt{2\alpha/\omega}$ and  
$\Delta_y \equiv \sqrt{2 \beta /\omega_0}$.
Making these displacements and then taking the thermodynamic limit results 
in a Hamiltonian $H^{(2)}_\mathrm{sc}$ that is identical with the 
quantum Hamiltonian $H^{(2)}$ of Eq. (\ref{hcDHab})
in the appropriate position-momentum 
representation, which  may thus be diagonalised with the same rotation.
The equilibrium positions of $H^{(2)}_\mathrm{sc}$ are 
$\rb{+\Delta_x,-\Delta_y}$ 
and $\rb{-\Delta_x,+\Delta_y}$.

The bounds on the existence of these classical effective Hamiltonians 
are exactly as in the quantum case -  
the excitation energies $\varepsilon_-^{(1)}$ and 
$\varepsilon_-^{(2)}$ of the decoupled modes 
remain real only on their respective sides of the critical 
coupling $\lambda_c$, which has the same value as in the quantum case.
Clearly the semi-classical system is completely integrable in this 
thermodynamic limit.

\subsection{Equations of motion}
To analyse the behaviour of this semi-classical system for finite $j$, 
we form Hamilton's equations of motion from 
the derivatives of $H_\mathrm{sc}$ \cite{gold}
\begin{eqnarray}
\dot{x} &=& p_x
\nonumber\label{xdot}\\
\dot{y} &=& p_y
         \rb{
	       1 - \frac{\lambda}{2 j}\sqrt{\frac{\omega}{\omega_0}} 
	             \frac{x y}{\sqrt{1 -\eta}}
            }
\nonumber\label{ydot}\\
\dot{p_x} &=& - \omega^2 x - 2 \lambda \sqrt{\omega \omega_0} 
	                         ~y \sqrt{1-\eta}
\nonumber\label{pxdot}\\
\dot{p_y} &=& -\omega_0^2 y - 2\lambda \sqrt{\omega \omega_0} ~x \sqrt{1-\eta}
	                       \rb{
				   1 - \frac{\omega_0 y^2}{4j\rb{1-\eta}}
				  }
\label{pydot}
\end{eqnarray}
where as before,
\begin{eqnarray}
\eta = \frac{1}{4 j \omega_0}\rb{\omega_0^2 y^2 + p_y^2 - \omega_0}.
\end{eqnarray}
It is not {\it a priori} obvious that this flow should preserve the 
condition set out in Eq. (\ref{etacond}).
However, we have demonstrated numerically that, 
providing we choose initial 
conditions that satisfy Eq. (\ref{etacond}), then this condition is 
always satisfied.  Although we have not shown this analytically, 
it can at least be seen to be plausible.  Calculating 
$\dot{\eta}=d\eta/dt= \left\{ H,\eta\right\}$ where 
$\left\{ \ldots \right\}$ denote Poisson brackets, we find that
\begin{eqnarray}
\dot{\eta} =
-\frac{\lambda}{j}\sqrt{\frac{\omega}{\omega_0}} x p_y \sqrt{1-\eta},
\end{eqnarray}
so that as $\eta$ approaches unity its rate of change approaches zero, 
implying that it is bound appropriately.

We now determine the fixed points of this flow at finite $j$ 
by setting 
$\dot{x}=\dot{y}=0$, $\dot{p_x} =\dot{p_y}=0$.   The simplest fixed point 
is given by $x=y=$$p_x=p_y=0$, the co-ordinate origin.  
By calculating the Hessian stability matrix from the 
second derivatives of $H$, we see that this fixed point is only stable 
when
\begin{eqnarray}
\frac{1}{2}
\left\{
  \omega^2 + \omega_0^2 
  - \sqrt{
          \rb{\omega^2 - \omega_0^2} 
          + 16 \lambda^2 \omega \omega_0\rb{1 + 1/\rb{4j}}
         }
\right\}
>0.
\end{eqnarray}
i.e. when
\begin{eqnarray}
\lambda < \frac{\lambda_c}{\sqrt{1 + 1/\rb{4j}}}.
\end{eqnarray}
There are two other fixed points, both of which have $p_x=p_y=0$, and 
with $x$ and $y$ given by
\begin{eqnarray}
x_0 = \pm \frac{2\lambda}{\omega}
	  \sqrt{ \frac{j}{\omega} 
                \left\{
                      \rb{1 +\frac{1}{4j}}^2 -\frac{\lambda_c^4}{\lambda^4}  
	       \right\}
	       }
&;&
y_0= \mp \sqrt{\frac{2j}{\omega} 
			\rb{
		            1 + \frac{1}{4j} 
			    - \frac{\lambda_c^2}{\lambda^2}
			   }
            }.
\end{eqnarray}
These two quantities only remain real provided that 
$1 + \frac{1}{4j}- \frac{\lambda_c^2}{\lambda^2} > 0$, 
which corresponds to the condition
\begin{eqnarray}
\lambda > \frac{\lambda_c}{\sqrt{1 + 1/\rb{4j}}}.
\end{eqnarray}
Provided that the above condition is fulfilled, the fixed points 
given by $\rb{+x_0,-y_0}$ and  $\rb{-x_0,+y_0}$ exist and are stable.
%%%%%%%%%%%%%%%%%%%%%%%%%%%%%%%%%%%%%%%%%%%%%%%%%%%%%%%%%%%%%%%%%%%
%%%%%%%%%%%%%%%%%%%%%%%%%%%%%%%%%%%%%%%%%%%%%%%%%%%%%%%%%%%%%%%%%%%
\begin{figure}[tb]
  \centerline{
    \includegraphics[clip=true,width=0.3\textwidth]{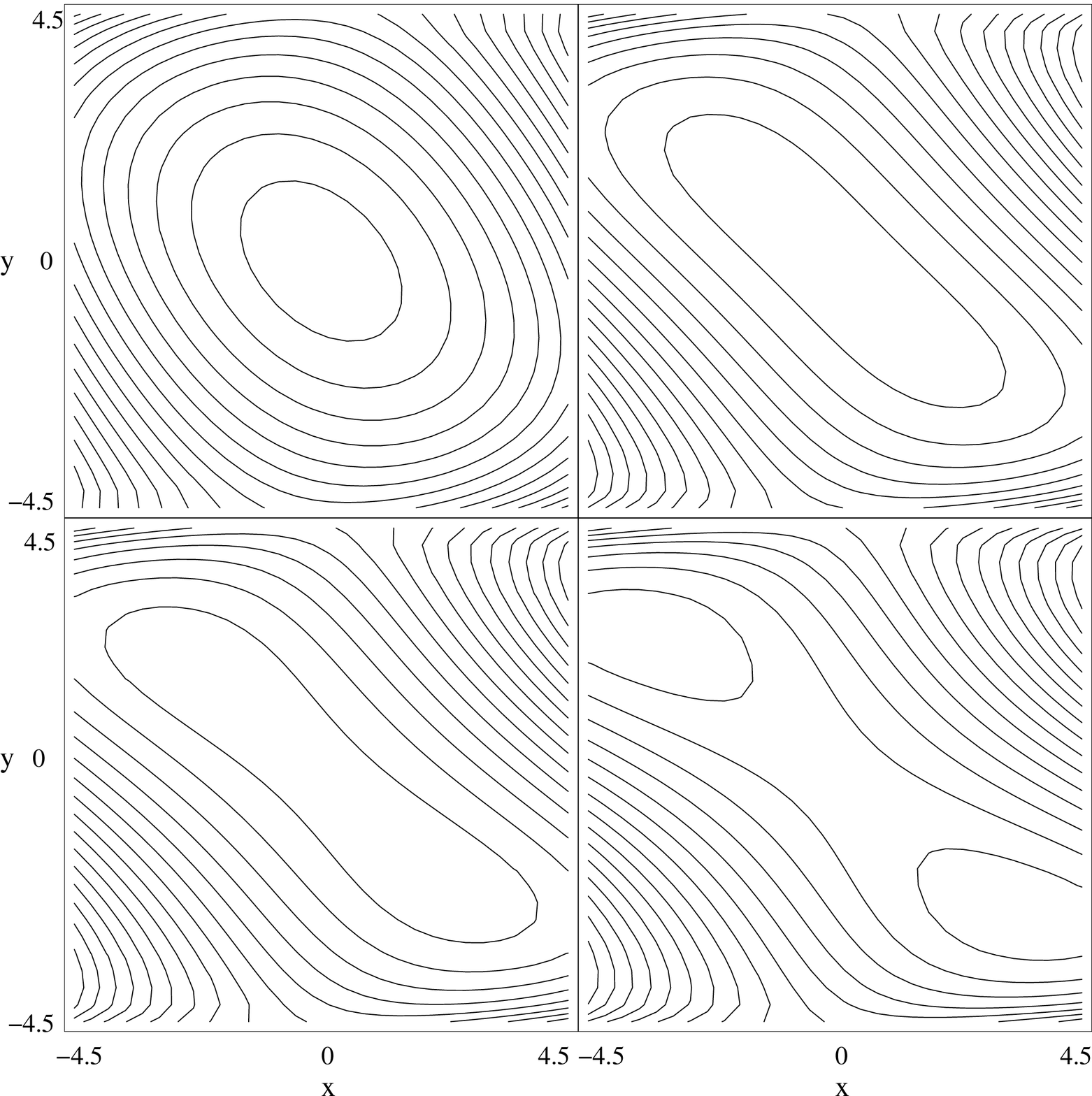}
    ~~~~
    \includegraphics[clip=true,width=0.3\textwidth]{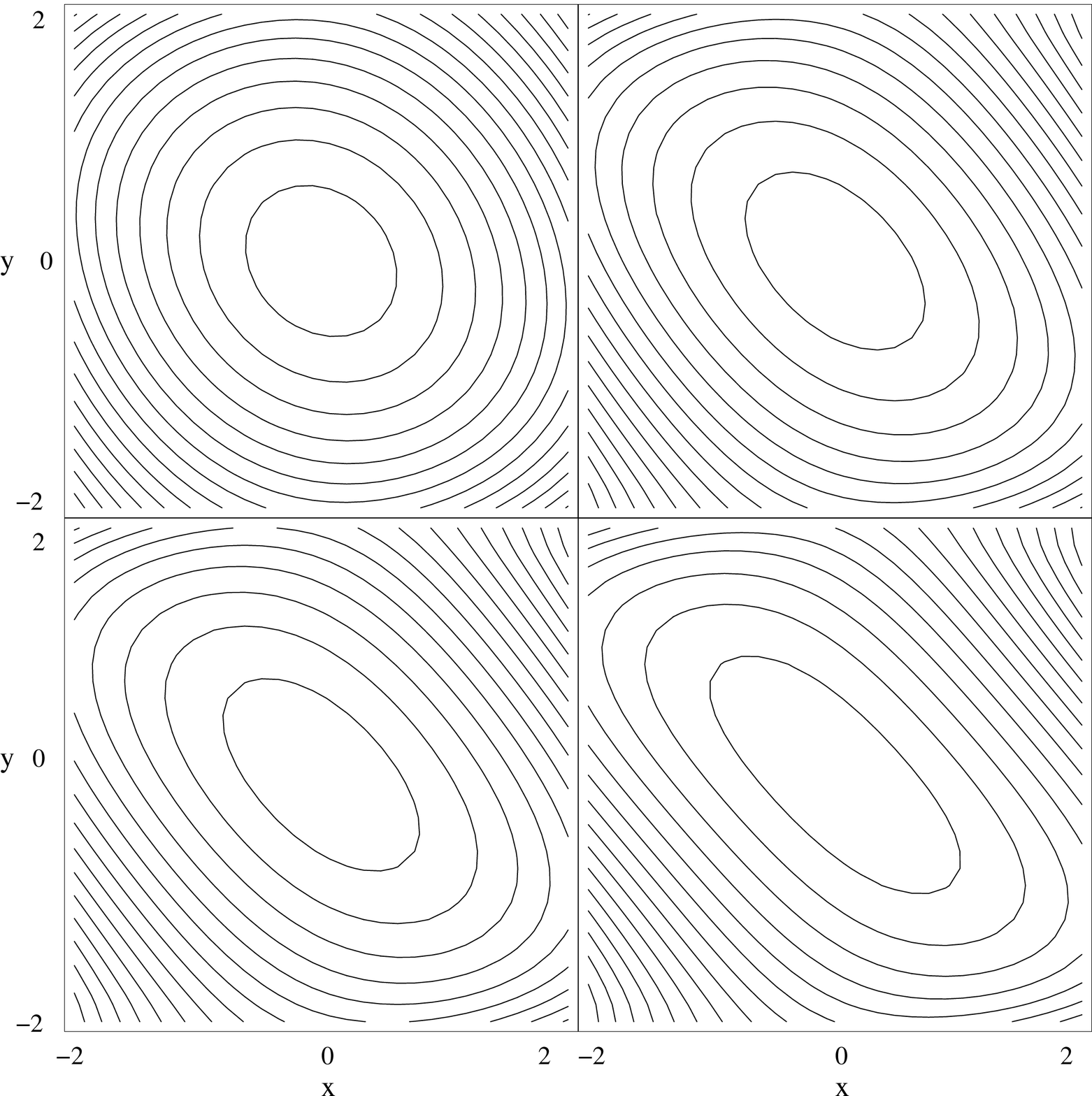}
  }
  \caption{\label{Upy}
    The momentum-dependent potential $U\rb{x,y,p_y}$ at two different values of
    momentum $p_y=0$ (left) and $p_y=3$ (right), for a series of 
    couplings the same as in
    Fig. (\ref{wavefn}).  Note the difference in scales between the two 
    plots.
    The Hamiltonian is resonant: $\omega = \omega_0 = 1$, 
    $\lambda_c=0.5$
  }
\end{figure}
%%%%%%%%%%%%%%%%%%%%%%%%%%%%%%%%%%%%%%%%%%%%%%%%%%%%%%%%%%%%%%%%%%
%%%%%%%%%%%%%%%%%%%%%%%%%%%%%%%%%%%%%%%%%%%%%%%%%%%%%%%%%%%%%%%%%%
So, below the coupling $\lambda_c / \sqrt{1 + 1/\rb{4j}}$, 
only one fixed 
point exists, which lies at the co-ordinate origin and is stable.  Above 
$\lambda =\lambda_c / \sqrt{1 + 1/\rb{4j}}$, this fixed point becomes
unstable and two new stable fixed points appear at the coordinates 
$\rb{+x_0,-y_0}$ and  $\rb{-x_0,+y_0}$.  Note that these expressions
give us the 
first correction to the location of the critical coupling 
in terms of a perturbation series in $j$.

We can consider this semi-classical system as a particle moving in the 
two-dimensional, momentum-dependent potential
\begin{eqnarray}
U\rb{x,y,p_y}= 
   \frac{1}{2} 
   \rb{
        \omega^2 x^2 +  \omega_0^2 y^2
      }
   + 2 \lambda \sqrt{
                        \omega \omega_0} 
                        ~x y \sqrt{1 -\frac{
                                          \omega_0^2 y^2 
                                           + p_y^2 - \omega_0}
                                           {4 j \omega_0}
                       }
  .
\end{eqnarray}
Maps of this potential for different values of increasing coupling and for 
two different values of $p_y$ are shown in Fig. \ref{Upy}.
Firstly, note how greatly the value of $p_y$ affects the shape 
of the potential felt by the ``particle''. For example, above $\lambda_c$ 
at $\lambda=0.8$, with $p_y=0$ the potential bifurcates
into two separate wells, whereas for $p_y=3$ it does not.
Also note the similarity between the plot of $U\rb{x,y,p_y}$ 
for $p_y=0$ and the plot of the wavefunction in Fig. \ref{wavefn}.  It 
is clear that the $p_y=0$ potential largely determines the structure of the 
wavefunction at finite $j$, presumably because the location of the 
fixed points are determined with $p_y=0$.

\subsection{Chaos in the Semi-Classical model}
\begin{figure}[tb]
  \centerline{
%    \includegraphics[clip=true,width=0.7\textwidth]
%      {./prettypoincarej5l_0_2and0_6.eps}
%  }
   \includegraphics[clip=true,width=0.7\textwidth]
      {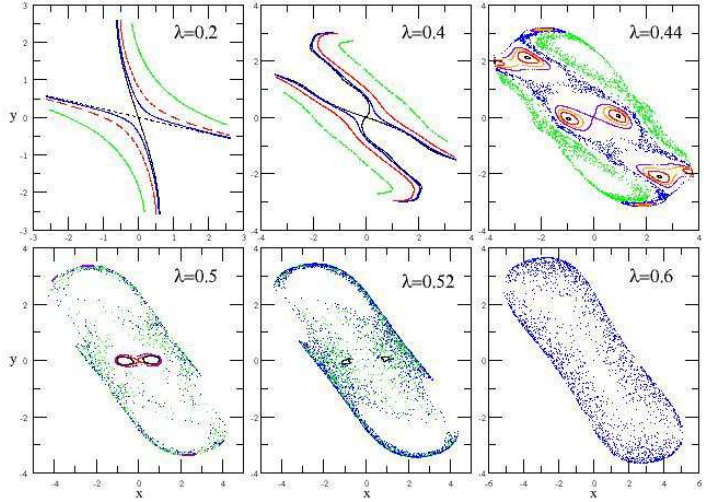}
  }
  \caption{\label{Psec}
    Poincare sections for the classical Dicke Model for a sequence 
    of increasing couplings, with $j=5$ and $E=-3$. 
    The Hamiltonian is resonant $\omega = \omega_0 = 1$;$\lambda_c=0.5$ 
  }
\end{figure}

We numerically integrate Hamilton's equations of motion
for the semi-classical system for a variety of different 
parameters and initial 
conditions.  In order to analyse the trajectories resulting from these
integrations, we use Poincar\'e sections through the four-dimensional 
phase-space.  As this system is Hamiltonian, the energy,
\begin{eqnarray}
 E = -j \omega_0 
       + \frac{1}{2} 
          \rb{
        \omega^2 x^2 + p_x^2 - \omega +  \omega_0^2 y^2 + p_y^2 - \omega_0
             }
    + 2 \lambda \sqrt{\omega \omega_0} ~x y \sqrt{1 - \eta}
\end{eqnarray}
is conserved, and thus we define our surface of section by $p_x=0$ 
with $p_y$ being fixed by the energy $E$.  We only record traversals 
for $p_y >0$.
Poincare sections for illustrative parameter values are shown in
Fig. \ref{Psec}.

At low $\lambda$ ($\lambda \le 0.4$ in Fig. \ref{Psec}), the Poincar\'e
sections consist of a series of regular, periodic orbits.  
Approaching the critical
coupling ($\lambda=0.44, 0.5$ in Fig. \ref{Psec}), we see a change in
the character of the periodic orbits and also the emergence of 
a number of chaotic trajectories.  Increasing the coupling further 
results in the break up of the remaining periodic orbits and the 
whole phase space becomes chaotic for couplings a little over the 
critical value ($\lambda=0.6$ in Fig. \ref{Psec}).  This transition to 
chaos in the classical system mirrors very closely that seen in the 
quantum system, especially in the way that most of the change in 
the nature of the behaviour is centred about the critical coupling 
determined by the phase-transition.
%%%%%%%%%%%%%%%%%%%%%%%%%%%%%%%%%%%%%%%%%%%%%%%%%%%%%%%%%%%%%%%%%%%
%%%%%%%%%%%%%%%%%%%%%%%%%%%%%%%%%%%%%%%%%%%%%%%%%%%%%%%%%%%%%%%%%%

An interesting feature of this classical Hamiltonian 
is that the (re-)quantisation of this Hamiltonian is not unique.  This is 
because the potential $U\rb{x,y,p_y}$ depends on the momentum $p_y$,  
a situation which may be compared to the quantisation of a Lagrangian 
for an electron 
in a magnetic field, where an extra `rule' is required to obtain the 
correct quantisation.
We may requantise $H_\mathrm{sc} $ by simply reversing the
steps in Eqs (\ref{clDHam1}-\ref{Hcl}).  However this is not the most obvious 
path as it involves the addition of extra imaginary $p_y$-dependent 
terms which have canceled in the final Hamiltonian.  Alternatively, one may
simply requantise Eq. (\ref{Hcl}) as it stands, 
which results in the Hamiltonian
\begin{eqnarray}
H' = \omega_0 \rb{b^\dagger b - j} + \omega a^\dagger a
+ \lambda \rb{a^\dagger + a}
\rb{ b^\dag + b} \sqrt{1-\frac{b^\dagger b}{2j}}  ,
\label{H'}
\end{eqnarray}
which is clearly different to the original bosonic Hamiltonian of
Eq. (\ref{clDHam1}).
This ambiguity disappears in the thermodynamic limit as here $U\rb{x,y,p_y}$
becomes  momentum independent in this limit in both of the systems phases. 

We note that the classical Hamiltonian
\begin{eqnarray}
H''= -j \omega_0 
   &+& \frac{1}{2} 
   \rb{
        \omega^2 x^2 + p_x^2 - \omega +  \omega_0^2 y^2 + p_y^2 - \omega_0
      }
\nonumber \\
    &+& 2 \lambda \sqrt{
                        \omega \omega_0} 
                        ~x y \sqrt{1 -\frac{
                                          \omega_0^2 y^2 
                                          - \omega_0}
                                           {4 j \omega_0}}
\end{eqnarray}
which is the same as the original Hamiltonian of Eq. (\ref{Hcl}) but with
$p_y^2$ removed from the square root, displays similar behaviour to that
of the full Hamiltonian.  The gain in simplicity in using this model, 
suggests that it would be an ideal test model for further exploration 
of the dynamics of this type of Hamiltonian constrained by a square-root.
The behaviour of the Hamiltonian  $H''$ and fact that the 
$p_y =0$ potential largely dominates dynamics of $H_{cl}$ and the 
structure of the wavefunction of the original DH, suggest that 
the requantisation route is not critical provided that $j$ is not small.

% LocalWords:  Dicke Hartree Fock dt wavefunction Poincar requantisation DH al
% LocalWords:  Primakoff Eq commutators QPT Hamiltonians cl decoupled co priori
% LocalWords:  ordinates diagonalised integrable behaviour ordinate requantise
% LocalWords:  quantisation bosonic photonic chaoticity Furuya sc

\newpage
\section{The RWA and Integrability \label{RWA}}
  
The DH in the RWA is given by
\begin{eqnarray}
H_\mathrm{RWA} = \omega_0 J_z + \omega a^\dagger a 
     + \frac{\lambda}{\sqrt{2j}} \rb{a^\dagger J_- + a J_+}. 
\label{DHamRWA}
\end{eqnarray}
It is in this form that the DH is generally studied and in which 
the thermodynamics of the phase transition were
originally discussed \cite{he:li,wa:ho}.  In the
RWA, the QPT occurs at a coupling twice that of the non-RWA critical 
value $ \lambda_c^{\mathrm{RWA}} = 2 \lambda_c = \sqrt{\omega \omega_0}$ 
\cite{he:li2,cr:du}.  
This is simply a consequence of the fact that in the non-RWA DH there 
are four terms in the interaction, whereas here we only have two.  As 
each term contributes to the mean-field, the
critical coupling of the RWA is twice as big as the non-RWA one.

In the RWA, the excitation number $\hat{N}$ of 
Eq. (\ref{parity}) becomes exactly conserved.
This splits the total Hilbert space into an infinite number 
of sub-spaces, labeled by the excitation number $\hat{n} = 0,1,2,\ldots$,
which in turn leads to level crossings and to a Poisson distribution for 
the $P\rb{S}$.
The crossover between the RWA and non-RWA $P\rb{S}$ distributions 
has been studied by treating the non-RWA terms as a perturbation 
\cite{le:ne}, and it was found that as the strength of this 
perturbation is increased from zero to one, 
a standard crossover between Poissonian and Wigner-Dyson statistics
is observed.

Here we wish to report two observations concerning 
the difference between the RWA and non-RWA models.
Firstly; a calculational issue that arises when considering the 
RWA system in the thermodynamic limit.  We may derive effective 
Hamiltonians in each phase, by using the Holstein-Primakoff 
representation as before.  In the normal phase, we  obtain
\begin{eqnarray}
H^{(1)}_\mathrm{RWA} = \omega_0 b^\dagger b + \omega a^\dag a 
		     + \lambda \rb{a^\dag b + b^\dag a} - j \omega_0.
\end{eqnarray}
The Bogoliubov transformations required to diagonalise this 
Hamiltonian are much simpler in terms of annihilation and creation operators 
than those for the non-RWA case.  
Specifically, the RWA diagonalising transformations are
\begin{eqnarray}
  a \go -c_1\sin \beta +  c_2\cos \beta
  ;~~~
  b \go c_1 \cos \beta + c_2\sin\beta ,
\end{eqnarray}
plus the Hermitian conjugate relations, where the rotation angle $\beta$ 
is given by  
\begin{eqnarray}
\tan\rb{2\beta} = \frac{2 \lambda}{\omega - \omega_0}.
\end{eqnarray}

The transformation for annihilation operators only 
involves annihilation operators, and the same with the creation operators.
This is to be contrasted with the non-RWA transformations, which 
transform any given operator into a linear combination of 
all four operators.
Therefore, in the RWA it is 
very simple to find the diagonalising transformation in the 
second quantised representation, where as in the non-RWA case, 
this diagonalisation only becomes transparent when one considers 
the first quantised position-momentum representation of the operators. 
The converse of this statement is true; it is hard to find the
diagonalising transformation in the RWA if one works in the
position-momentum representation.  We conjecture that this is a more 
general point than just applying here, and hope that this observation 
may be useful in other problems. 

\begin{figure}[tb]
%\centerline{\includegraphics[clip=true,width=0.4\textwidth]
%            {./spectrumj5.eps}
%            ~~~~
%            \includegraphics[clip=true,width=0.4\textwidth]
%             {./RWAspectrumj5.eps}
%            }
\centerline{\includegraphics[clip=true,width=0.4\textwidth]
            {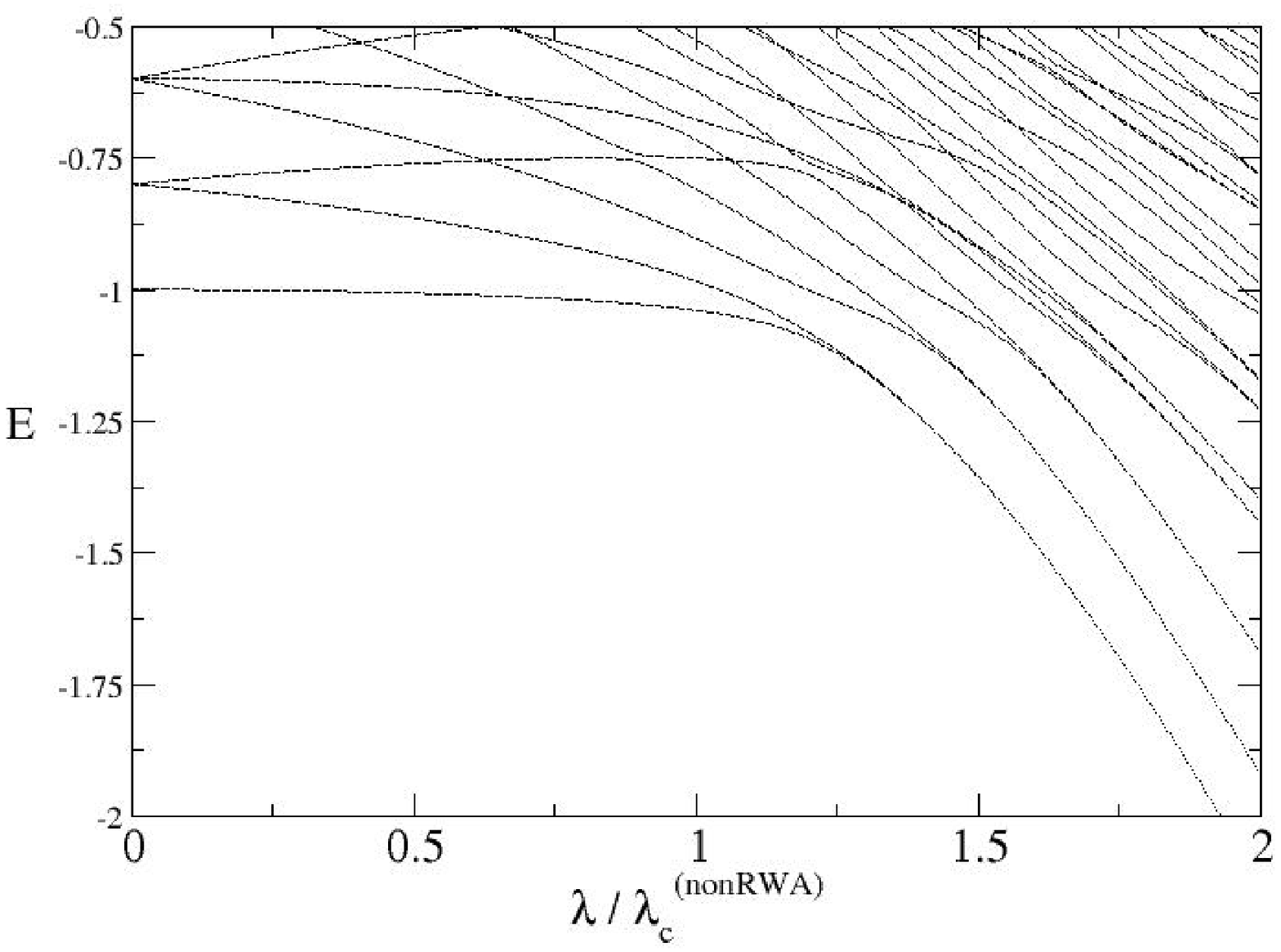}
            ~~~~
            \includegraphics[clip=true,width=0.4\textwidth]
             {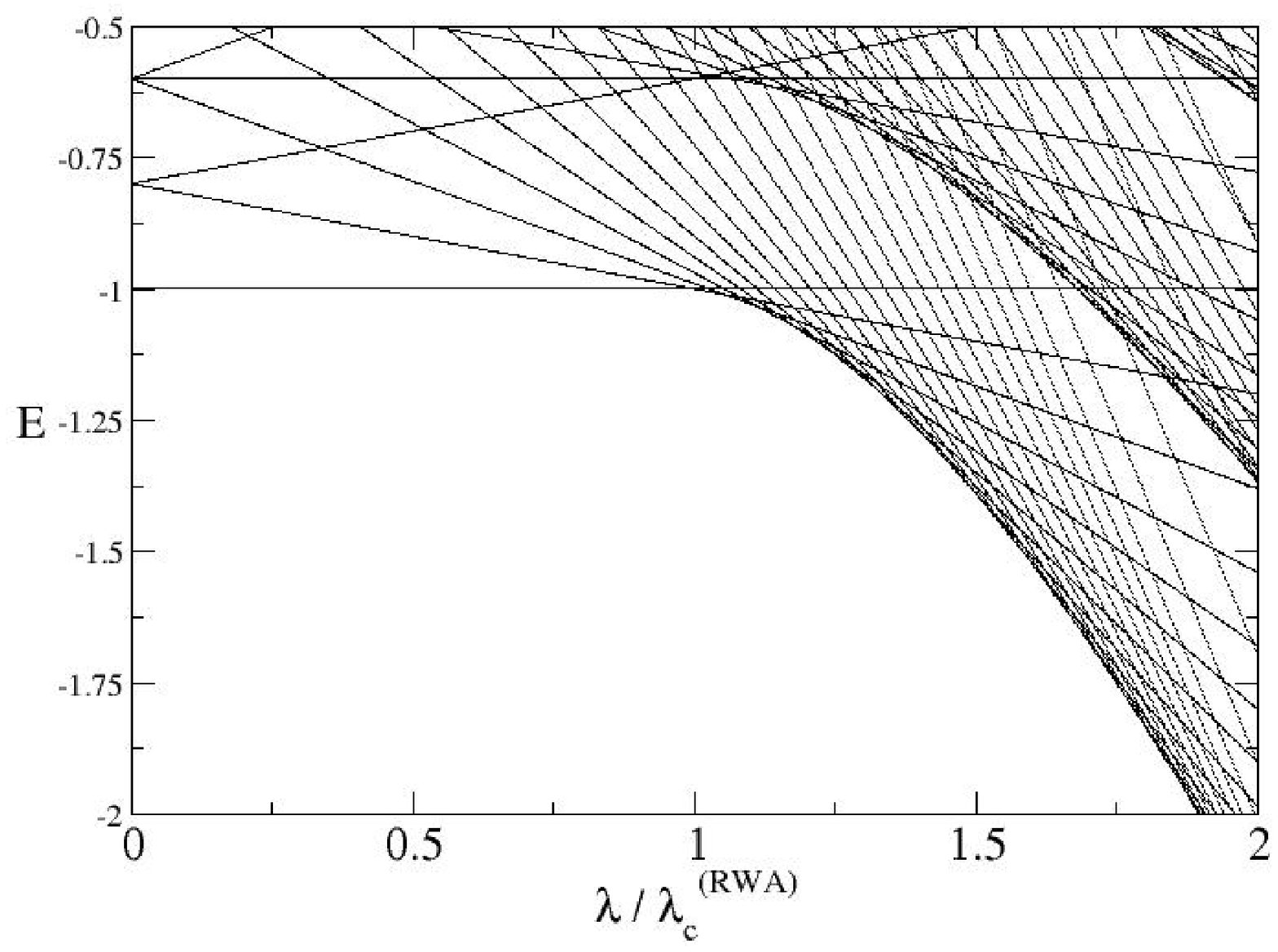}
            }
\caption{\label{rwanonrwa}
The full energy schema of the (a) non-RWA and (b) RWA Dicke Hamiltonian 
for j=5.
The Hamiltonian is resonant; $\omega = \omega_0 = 1$.  }
\end{figure}

Our second observation concerns the comparison of the energy spectra 
at finite $j$ of the RWA \cite{na:or} 
and non-RWA Hamiltonians.  Figure \ref{rwanonrwa}
shows two typical spectra, with coupling axes chosen for easy comparison.
%%%%%%%%%%%%%%%%%%%%%%%%%%%%%%%%%%%%%%%%%%%%%%%%%%%%%%%%%%%%%%%%%%%
%%%%%%%%%%%%%%%%%%%%%%%%%%%%%%%%%%%%%%%%%%%%%%%%%%%%%%%%%%%%%%%%%%%
%%%%%%%%%%%%%%%%%%%%%%%%%%%%%%%%%%%%%%%%%%%%%%%%%%%%%%%%%%%%%%%%%%%
%%%%%%%%%%%%%%%%%%%%%%%%%%%%%%%%%%%%%%%%%%%%%%%%%%%%%%%%%%%%%%%%%%%
In terms of the appropriate critical coupling, the 
ground-state energy of the non-RWA spectrum is remarkably well 
approximated by the caustic of all the energy levels in the RWA 
spectrum that have negative slopes.  As $j$ increases this approximation 
becomes better, as the length of the individual line segments become 
shorter until, in the thermodynamic limit, the correspondence 
of the ground-states 
becomes exact and both excitation spectra become quasi-continuous.

% LocalWords:  DH RWA superradiance QPT  integrability Eq Poissonian
% LocalWords:  calculational Hamiltonians Primakoff Bogoliubov diagonalise
% LocalWords:  diagonalising Hermitian quantised diagonalisation Dicke Dyson

\section{Discussion \label{disc}}
  
We have presented a coherent and comprehensive picture of how 
the existence of a QPT in the thermodynamic limit plays a 
crucial role in determining chaotic properties in a 
model interacting system.  The DH exhibits a change-over 
from quasi-integrability to chaos, and this transition 
is located by the precursors of the QPT around the 
critical point, $\lambda_c$.  This statement applies 
equally well to the original quantum system and to 
the semi-classical counterpart derived from it.

Our analysis of the DH in the thermodynamic limit consists of 
deriving an effective Hamiltonian to describe the system in
each of its normal and super-radiant phases.  For arbitrary 
coupling the system is described in terms of two decoupled
modes, each of which is a collective photon-atom excitation, and it
is the vanishing of the excitation energy associated with 
the photon-like mode that delimits the two phases.
Our approach is particularly useful because we can calculate
exactly any property of the system in the thermodynamic limit, 
by simply utilising the appropriate Bogoliubov
transformations.

This analysis reveals that the QPT breaks
the symmetry associated with the parity operator $\Pi$.  
In the normal phase, where the system in effectively unexcited,
the wavefunctions of the system are invariant with respect to $\Pi$.
In the super-radiant phase however, this global symmetry is 
broken and two new local symmetries appear, 
each of which describes an isolated 
wavefunction lobe, and the spectrum is doubly degenerate.
This symmetry breaking strictly only occurs 
in the thermodynamic limit,
and at any finite j, these lobes are joined together in 
a total wavefunction that is $\Pi$-invariant.
That these two lobes are separated 
by a macroscopic amount, 
proportional to the square-root of the system size, means 
the onset of chaos is accompanied by the delocalisation of the 
wavefunctions and the appearance of macroscopic coherence 
in the system.

Similar features occur in the three-dimensional 
Anderson model.  This model of a disordered 
electron system exhibits a metal-insulator 
transition, in which the wavefunctions 
are localised for strong disorder and delocalised when the 
disorder is weak.
\cite{le:ra,zh:kr,prange}.  
Analysis of the level-statistics 
shows that the $P\rb{S}$ changes from Poissonian 
to Wigner-Dyson at the phase-transition point, which is 
determined by the magnitude of the random potential 
fluctuations.
It is remarkable that our comparatively simple model 
should bear so many important features in common with 
complex disorder models, such as the Anderson model,
although one feature of such models that we have found 
no evidence of in the DH is the existence of a third 
universal $P\rb{S}$ distribution precisely at the 
critical coupling \cite{zh:kr}.

There are two different classical limits involved with 
the Dicke model, and by extension, models of similar
nature involving quantum spins and boson fields.  
Firstly there is the limit of $j \go \infty$, in which 
the length of the spin becomes macroscopic.  The second 
is the limit $\hbar \go 0$, which we have performed here 
when setting bosonic commutators equal to zero. 

These limits may be applied independently and in 
either order.  If we apply the $j \go \infty$ limit 
first to the DH, we obtain the effective Hamiltonians 
$H^{(1,2)}$.  Taking then $\hbar \go 0$ by setting the
commutators of the collective modes to zero, we 
simply obtain $H^{(1,2)}_\mathrm{sc}$, the two
classical effective Hamiltonians.  Note that the
integrability of $H^{(1,2)}$ makes this 
``de-quantisation''  direct and unambiguous.
Applying this limit in the other order means that
starting with the DH in the Holstein-Primakoff
representation we set $\hbar \go 0$ by setting 
the original field and atom bosonic commutators 
to zero, which results in our semi-classical 
Hamiltonian $H_\mathrm{sc}$.  Subsequent taking 
of the $j \go \infty$ limit results in 
$H^{(1,2)}_\mathrm{sc}$ as above, showing that
we obtain the same result independently of the 
order in which the limits are taken.

After both limits, the system described by 
$H^{(1,2)}_\mathrm{sc}$ is ``the classical'' 
analogue of the DH, describing a macroscopic collection 
of atoms in terms of classical variables.  This system 
is completely integrable, and there is no sign of chaos 
either in it or its quantised counterpart $H^{(1,2)}$.

These results support the recent argument put forward by Ballentine 
concerning the existence of so-called ``semi-quantum chaos'' \cite{ball}. 
Semi-quantum chaos is that which arises from the coupling of a
quantum and a classical system, neither of which are by
themselves chaotic.  Ballentine studied a model of a massive
particle of mass $m$ interacting with a spin-half. By 
considering the semi-classical limit of $m \go \infty$, the 
semi-quantum system of a quantum spin interacting with a classical
particle was realised.  He demonstrated that as
$m \go \infty$, the chaos in the system rapidly disappeared.  Our 
results here may be seen as the complement to this system, where
the mass is kept constant but the length of the pseudo-spin is
taken to the classical limit $j \go \infty$.  
Given the integrability of the DH is this limit, 
there is certainly no semi-quantum chaos in our system, which 
lends additional weight to Ballentine's claim that 
semi-quantum chaos does not exist.

The question then arises what is the status of the two
systems obtained by only taking one of the two limits.  
In the case of only taking the $j \go\infty$ limit, the 
answer is simple; $H^{(1,2)}$ is a direct quantisation of 
$H^{(1,2)}_\mathrm{sc}$ and describes quantum fluctuations
around classical mean-fields.  More interesting is the 
status of $H_\mathrm{sc}$.  We have shown here that its behaviour 
matches very closely that of the quantum DH, and that it
has been derived in an almost canonical way, so its
mathematical status as the semi-classical counterpart of 
the DH seems reasonably secure, but what the relevance of 
this model to the physical system is less obvious.  

The nature of the $\hbar \go 0$ limit suggests that
this model might
be useful in describing the model when there are a few 
atoms (10-20) present, and almost-classical fields, i.e. 
coherent states, 
are applied.  Under these circumstances the original DH 
and semi-classical model $H_\mathrm{sc}$ might be fruitfully 
compared.

% LocalWords:  QPT integrability DH decoupled Bogoliubov wavefunction realised
% LocalWords:  Ballentine behaviour Dicke delocalisation decoherence utilising
% LocalWords:  superpositions unexcited wavefunctions delocalised Poissonian sc
% LocalWords:  Dyson boson bosonic commutators Hamiltonians de quantisation
% LocalWords:  Primakoff integrable quantised Ballentine's

\begin{acknowledgments}
  This work was supported by projects EPSRC GR44690/01, DFG Br1528/4-1,
  the WE Heraeus foundation and the UK Quantum Circuits Network. 
\end{acknowledgments}

%%%%%%%%%%%%%%%%%%%%%%%%%%%%%%%%%%%%%%%%%%%%%%%%%%%%%%%%%%%%%%%%%%%
\newpage
\begin{appendix}
\section{Bogoliubov Transformation \label{appBOGT}}
\subsection{Normal phase}
The two sets of bosons, $\left\{a,b\right\}$ and 
$\left\{c_1,c_2\right\}$, may be expressed in terms of
one another as
\begin{eqnarray}
a^\dagger &=& \frac{1}{2}
  \left\{  
    \frac{\cos \gamma^{(1)}}{\sqrt{\omega\varepsilon^{(1)}_-}}
    \left[
      \rb{\omega + \varepsilon^{(1)}_-}c_1^\dagger +  \rb{\omega - \varepsilon^{(1)}_-} c_1
    \right]
    + \frac{\sin \gamma^{(1)}}{\sqrt{\omega\varepsilon^{(1)}_+}}
    \left[
      \rb{\omega + \varepsilon^{(1)}_+}c_2^\dagger +  \rb{\omega - \varepsilon^{(1)}_+} c_2
    \right]
  \right\}, \nonumber \\
a  &=&  \frac{1}{2}
  \left\{  
    \frac{\cos \gamma^{(1)}}{\sqrt{\omega\varepsilon^{(1)}_-}}
    \left[
      \rb{\omega - \varepsilon^{(1)}_-}c_1^\dagger +  \rb{\omega + \varepsilon^{(1)}_-} c_1
    \right]
    + \frac{\sin \gamma^{(1)}}{\sqrt{\omega\varepsilon^{(1)}_+}}
    \left[
      \rb{\omega - \varepsilon^{(1)}_+}c_2^\dagger +  \rb{\omega + \varepsilon^{(1)}_+} c_2
    \right]
  \right\}, \nonumber \\
b^\dagger &=&  \frac{1}{2}
  \left\{  
    \frac{-\sin \gamma^{(1)}}{\sqrt{\omega_0\varepsilon^{(1)}_-}}
    \left[
      \rb{\omega_0 + \varepsilon^{(1)}_-}c_1^\dagger +  \rb{\omega_0 - \varepsilon^{(1)}_-} c_1
    \right]
    + \frac{\cos \gamma^{(1)}}{\sqrt{\omega_0\varepsilon^{(1)}_+}}
    \left[
      \rb{\omega_0 + \varepsilon^{(1)}_+}c_2^\dagger +  \rb{\omega_0 - \varepsilon^{(1)}_+} c_2
    \right]
  \right\}, \nonumber \\
b &=&  \frac{1}{2}
  \left\{  
    \frac{-\sin \gamma^{(1)}}{\sqrt{\omega_0\varepsilon^{(1)}_-}}
    \left[
      \rb{\omega_0 - \varepsilon^{(1)}_-}c_1^\dagger +  \rb{\omega_0 + \varepsilon^{(1)}_-} c_1
    \right]
    + \frac{\cos \gamma^{(1)}}{\sqrt{\omega_0\varepsilon^{(1)}_+}}
    \left[
      \rb{\omega_0 - \varepsilon^{(1)}_+}c_2^\dagger +  \rb{\omega_0 + \varepsilon^{(1)}_+} c_2
    \right]
  \right\},\label{lcbog1}
\end{eqnarray}
with the inverse relations
\begin{eqnarray}
c_1^\dagger &=&\frac{1}{2}
  \left\{  
    \frac{\cos \gamma^{(1)}}{\sqrt{\omega\varepsilon^{(1)}_-}}
    \left[
      \rb{\varepsilon^{(1)}_- + \omega}a^\dagger +  \rb{\varepsilon^{(1)}_- - \omega} a
    \right]
    - \frac{\sin \gamma^{(1)}}{\sqrt{\omega_0\varepsilon^{(1)}_-}}
    \left[
      \rb{\varepsilon^{(1)}_- + \omega_0} b^\dagger +  \rb{\varepsilon^{(1)}_- - \omega_0} b
    \right]
  \right\}, \nonumber \\
c_1 &=&\frac{1}{2}
  \left\{  
    \frac{\cos \gamma^{(1)}}{\sqrt{\omega\varepsilon^{(1)}_-}}
    \left[
      \rb{\varepsilon^{(1)}_- - \omega}a^\dagger +  \rb{\varepsilon^{(1)}_- + \omega} a
    \right]
    - \frac{\sin \gamma^{(1)}}{\sqrt{\omega_0\varepsilon^{(1)}_-}}
    \left[
      \rb{\varepsilon^{(1)}_- - \omega_0} b^\dagger +  \rb{\varepsilon^{(1)}_- + \omega_0} b
    \right]
  \right\}, \nonumber \\
c_2^\dagger &=&\frac{1}{2}
  \left\{  
    \frac{\sin \gamma^{(1)}}{\sqrt{\omega\varepsilon^{(1)}_+}}
    \left[
      \rb{\varepsilon^{(1)}_+ + \omega}a^\dagger +  \rb{\varepsilon^{(1)}_+ - \omega} a
    \right]
    + \frac{\cos \gamma^{(1)}}{\sqrt{\omega_0\varepsilon^{(1)}_+}}
    \left[
      \rb{\varepsilon^{(1)}_+ + \omega_0} b^\dagger +  \rb{\varepsilon^{(1)}_+ - \omega_0} b
    \right]
  \right\}, \nonumber \\
c_2 &=& \frac{1}{2}
  \left\{  
    \frac{\sin \gamma^{(1)}}{\sqrt{\omega\varepsilon^{(1)}_+}}
    \left[
      \rb{\varepsilon^{(1)}_+ - \omega}a^\dagger +  \rb{\varepsilon^{(1)}_+ + \omega} a
    \right]
    + \frac{\cos \gamma^{(1)}}{\sqrt{\omega_0\varepsilon^{(1)}_+}}
    \left[
      \rb{\varepsilon^{(1)}_+ - \omega_0} b^\dagger +  \rb{\varepsilon^{(1)}_+ + \omega_0} b
    \right]
  \right\}.
\end{eqnarray}
The angle $\gamma^{(1)}$ is rotation angle of the coordinate system
which eliminates the interaction in the position representation, and is
given by
\begin{equation}
\tan \rb{2\gamma^{(1)}} = \frac{4 \lambda \sqrt{\omega \omega_0}}
{\omega_0^2 - \omega^2}.
\end{equation}
%%%%%%%%%%%%%%%%%%%%%%%%%%%%%%%
\subsection{Super-radiant phase}
The analogous Bogoliubov transfomations is the super-radiant phase are
\begin{eqnarray}
c^\dagger &=& \frac{1}{2}
  \left\{  
    \frac{\cos \gamma^{(2)}}{\sqrt{\omega\varepsilon^{(2)}_-}}
    \left[
      \rb{\omega + \varepsilon^{(2)}_-}e_1^\dagger +  \rb{\omega - \varepsilon^{(2)}_-} e_1
    \right]
    + \frac{\sin \gamma^{(2)}}{\sqrt{\omega\varepsilon^{(2)}_+}}
    \left[
      \rb{\omega + \varepsilon^{(2)}_+}e_2^\dagger +  \rb{\omega - \varepsilon^{(2)}_+} e_2
    \right]
  \right\}, \nonumber \\
c  &=&  \frac{1}{2}
  \left\{  
    \frac{\cos \gamma^{(2)}}{\sqrt{\omega\varepsilon^{(2)}_-}}
    \left[
      \rb{\omega - \varepsilon^{(2)}_-}e_1^\dagger +  \rb{\omega + \varepsilon^{(2)}_-} e_1
    \right]
    + \frac{\sin \gamma^{(2)}}{\sqrt{\omega\varepsilon^{(2)}_+}}
    \left[
      \rb{\omega - \varepsilon^{(2)}_+}e_2^\dagger +  \rb{\omega + \varepsilon^{(2)}_+} e_2
    \right]
  \right\}, \nonumber \\
d^\dagger &=&  \frac{1}{2}
  \left\{  
    \frac{-\sin \gamma^{(2)}}{\sqrt{\widetilde{\omega}\varepsilon^{(2)}_-}}
    \left[
      \rb{\widetilde{\omega} + \varepsilon^{(2)}_-}e_1^\dagger +  \rb{\widetilde{\omega} - \varepsilon^{(2)}_-} e_1
    \right]
    + \frac{\cos \gamma^{(2)}}{\sqrt{\widetilde{\omega}\varepsilon^{(2)}_+}}
    \left[
      \rb{\widetilde{\omega} + \varepsilon^{(2)}_+}e_2^\dagger +  \rb{\widetilde{\omega} - \varepsilon^{(2)}_+} e_2
    \right]
  \right\}, \nonumber \\
d &=&  \frac{1}{2}
  \left\{  
    \frac{-\sin \gamma^{(2)}}{\sqrt{\widetilde{\omega}\varepsilon^{(2)}_-}}
    \left[
      \rb{\widetilde{\omega} - \varepsilon^{(2)}_-}e_1^\dagger +  \rb{\widetilde{\omega} + \varepsilon^{(2)}_-} e_1
    \right]
    + \frac{\cos \gamma^{(2)}}{\sqrt{\widetilde{\omega}\varepsilon^{(2)}_+}}
    \left[
      \rb{\widetilde{\omega} - \varepsilon^{(2)}_+}e_2^\dagger +  \rb{\widetilde{\omega} + \varepsilon^{(2)}_+} e_2
    \right]
  \right\}\label{hcbog1}
\end{eqnarray}
and
\begin{eqnarray}
e_1^\dagger &=&\frac{1}{2}
  \left\{  
    \frac{\cos \gamma^{(2)}}{\sqrt{\omega\varepsilon^{(2)}_-}}
    \left[
      \rb{\varepsilon^{(2)}_- + \omega}c^\dagger +  \rb{\varepsilon^{(2)}_- - \omega} c
    \right]
    - \frac{\sin \gamma^{(2)}}{\sqrt{\widetilde{\omega}\varepsilon^{(2)}_-}}
    \left[
      \rb{\varepsilon^{(2)}_- + \widetilde{\omega}} d^\dagger +  \rb{\varepsilon^{(2)}_- - \widetilde{\omega}} d
    \right]
  \right\}, \nonumber \\
e_1 &=&\frac{1}{2}
  \left\{  
    \frac{\cos \gamma^{(2)}}{\sqrt{\omega\varepsilon^{(2)}_-}}
    \left[
      \rb{\varepsilon^{(2)}_- - \omega}c^\dagger +  \rb{\varepsilon^{(2)}_- + \omega} c
    \right]
    - \frac{\sin \gamma^{(2)}}{\sqrt{\widetilde{\omega}\varepsilon^{(2)}_-}}
    \left[
      \rb{\varepsilon^{(2)}_- - \widetilde{\omega}} d^\dagger +  \rb{\varepsilon^{(2)}_- + \widetilde{\omega}} d
    \right]
  \right\}, \nonumber \\
e_2^\dagger &=&\frac{1}{2}
  \left\{  
    \frac{\sin \gamma^{(2)}}{\sqrt{\omega\varepsilon^{(2)}_+}}
    \left[
      \rb{\varepsilon^{(2)}_+ + \omega}c^\dagger +  \rb{\varepsilon^{(2)}_+ - \omega} c
    \right]
    + \frac{\cos \gamma^{(2)}}{\sqrt{\widetilde{\omega}\varepsilon^{(2)}_+}}
    \left[
      \rb{\varepsilon^{(2)}_+ + \widetilde{\omega}} d^\dagger +  \rb{\varepsilon^{(2)}_+ - \widetilde{\omega}} d
    \right]
  \right\}, \nonumber \\
e_2 &=& \frac{1}{2}
  \left\{  
    \frac{\sin \gamma^{(2)}}{\sqrt{\omega\varepsilon^{(2)}_+}}
    \left[
      \rb{\varepsilon^{(2)}_+ - \omega}c^\dagger +  \rb{\varepsilon^{(2)}_+ + \omega} c
    \right]
    + \frac{\cos \gamma^{(2)}}{\sqrt{\widetilde{\omega}\varepsilon^{(2)}_+}}
    \left[
      \rb{\varepsilon^{(2)}_+ - \widetilde{\omega}} d^\dagger +  \rb{\varepsilon^{(2)}_+ + \widetilde{\omega}} d
    \right]
  \right\}, \label{hcbog2}
\end{eqnarray}
where the angle $\gamma^{(2)}$ is given by
\begin{eqnarray}
\tan \rb{2 \gamma^{(2)}} = \frac{2\omega\omega_0\mu^2}
				{\omega_0^2 - \mu^2 \omega^2}
\end{eqnarray}
and where
\begin{eqnarray}
\tilde{\omega} \equiv \frac{\omega_0}{2}\rb{1+\frac{\lambda^2}{\lambda_c^2}}.
\end{eqnarray}

\section{Squeezing variances \label{sqzgvar}}
The preceeding Bogoliubov transformations may be used to derive
exact expressions for the squeezing
variances of the ground-state wavefunction in the thermodynamic 
limit as discussed in section \ref{TDL}.  In the normal phase they 
are given by
\begin{eqnarray}
  \rb{\Delta x }^2 &=&\frac{1}{2\omega}
    \rb{1 + 
      \frac{
       \varepsilon^{(1)}_+ \rb{\omega-\varepsilon^{(1)}_-}\cos^2\gamma^{(1)}
       +
       \varepsilon^{(1)}_- \rb{\omega-\varepsilon^{(1)}_+}\sin^2\gamma^{(1)} 
           }
           {
       \varepsilon^{(1)}_- \varepsilon^{(1)}_+
           }
}
\nonumber \\
\rb{\Delta p_x }^2 &=&\frac{\omega}{2}
    \rb{1 + 
      \frac{
       \rb{\varepsilon^{(1)}_- - \omega}\cos^2\gamma^{(1)}
       +
       \rb{\varepsilon^{(1)}_+ - \omega}\sin^2\gamma^{(1)} 
           }
           {
       \omega
           }
}
\end{eqnarray}
\begin{eqnarray}
 \rb{\Delta y }^2 &=&\frac{1}{2\omega_0}
    \rb{1 + 
      \frac{
       \varepsilon^{(1)}_+ \rb{\omega_0-\varepsilon^{(1)}_-}\sin^2\gamma^{(1)}
       +
       \varepsilon^{(1)}_- \rb{\omega_0-\varepsilon^{(1)}_+}\cos^2\gamma^{(1)} 
           }
           {
       \varepsilon^{(1)}_- \varepsilon^{(1)}_+
           }
}
\nonumber \\
\rb{\Delta p_y }^2 &=&\frac{\omega_0}{2}
    \rb{1 + 
      \frac{
       \rb{\varepsilon^{(1)}_- - \omega_0}\sin^2\gamma^{(1)}
       +
       \rb{\varepsilon^{(1)}_+ - \omega_0}\cos^2\gamma^{(1)} 
           }
           {
       \omega_0
           }
}
,
\end{eqnarray}
%%%%%%% %%%%%%%%%% %%%%%%%%%%%% %%%%%%%% %%%%%%%% %%%%%%%%% %%%%%%%% %%%%%%%
whereas in the super--radiant phase we find
\begin{eqnarray}
  \rb{\Delta x }^2 &=&\frac{1}{2\omega}
    \rb{1 + 
      \frac{
       \varepsilon^{(2)}_+ \rb{\omega-\varepsilon^{(2)}_-}\cos^2\gamma^{(2)}
       +
       \varepsilon^{(2)}_- \rb{\omega-\varepsilon^{(2)}_+}\sin^2\gamma^{(2)} 
           }
           {
       \varepsilon^{(2)}_- \varepsilon^{(2)}_+
           }
}
\nonumber \\
\rb{\Delta p_x }^2 &=&\frac{\omega}{2}
    \rb{1 + 
      \frac{
       \rb{\varepsilon^{(2)}_- - \omega}\cos^2\gamma^{(2)}
       +
       \rb{\varepsilon^{(2)}_+ - \omega}\sin^2\gamma^{(2)} 
           }
           {
       \omega
           }
}
\end{eqnarray}
\begin{eqnarray}
 \rb{\Delta y }^2 &=&\frac{1}{2\omega_0}
    \rb{1 + 
      \frac{
       \varepsilon^{(2)}_+ \rb{\tilde{\omega}
               -\varepsilon^{(2)}_-}\sin^2\gamma^{(2)}
       +
       \varepsilon^{(2)}_- \rb{\tilde{\omega}
               -\varepsilon^{(2)}_+}\cos^2\gamma^{(2)} 
           }
           {
       \varepsilon^{(2)}_- \varepsilon^{(2)}_+
           }
}
\nonumber \\
\rb{\Delta p_y }^2 &=&\frac{\omega_0}{2}
    \rb{1 + 
      \frac{
       \rb{\varepsilon^{(2)}_- - \tilde{\omega}}\sin^2\gamma^{(2)}
       +
       \rb{\varepsilon^{(2)}_+ - \tilde{\omega}}\cos^2\gamma^{(2)} 
           }
           {
       \tilde{\omega}
           }
}
.
\end{eqnarray}% LocalWords:  bosons Bogoliubov transfomations
These results are plotted in the main body of the text.

\end{appendix}

\end{document}